\documentclass[preprint,12pt]{elsarticle}

\usepackage{graphicx}
\usepackage{amsmath}
\usepackage{hyperref}
\hypersetup{
    colorlinks=true,       
}
\usepackage{subcaption}

\usepackage{amssymb}

\newcommand{\ii}{\mathrm{i}}
\newcommand{\e}{\mathrm{e}}
\newcommand{\del}{\boldsymbol{\nabla}}

\usepackage[]{algorithm2e}
\usepackage{pythonhighlight}
\usepackage{placeins}
\usepackage{nth}

\journal{Elsevier}

\begin{document}

\begin{frontmatter}

\title{Parallel-in-time integration of Kinematic Dynamos}

\author[1]{Andrew T. Clarke}
\ead{scatc@leeds.ac.uk}
\cortext[cor1]{Corresponding author: 
}
\author[2]{Christopher J. Davies}
\author[3]{Daniel Ruprecht}

\author[4]{Steven M. Tobias}

\address[1]{Centre for Doctoral Training in Fluid Dynamics, School of Computing, University of Leeds, Leeds, LS2 9JT, UK}

\address[2]{School of Earth and Environment, University of Leeds, Leeds, LS2 9JT, UK}
\address[3]{School of Mechanical Engineering, University of Leeds,  Leeds, LS2 9JT, UK}
\address[4]{Department of Applied Mathematics, University of Leeds, Leeds, LS2 9JT, UK}

\begin{abstract}

The precise mechanisms responsible for the natural dynamos in the Earth and Sun are still not fully understood. Numerical simulations of natural dynamos are extremely computationally intensive, and are carried out in parameter regimes many orders of magnitude away from real conditions.
Parallelization in space is a common strategy to speed up simulations on high performance computers, but eventually hits a scaling limit. 
Additional directions of parallelization are desirable to utilise the high number of processor cores now available. 
Parallel-in-time methods can deliver speed up in addition to that offered by spatial partitioning but have not yet been applied to dynamo simulations.
This paper investigates the feasibility of using the parallel-in-time algorithm Parareal to speed up initial value problem simulations of the kinematic dynamo, using the open source Dedalus spectral solver. 
Both the time independent Roberts and time dependent Galloway-Proctor 2.5D dynamos are investigated over a range of magnetic Reynolds numbers. 

Speed ups beyond those possible from spatial parallelisation are found in both cases. 
Results for the Galloway-Proctor flow are promising, with Parareal efficiency found to be close to 0.3. Roberts flow results are less efficient, but Parareal still shows some speed up over spatial parallelisation alone.

Parallel in space and time speed ups of $\sim300$ were found for 1600 cores for the Galloway-Proctor flow, with total parallel efficiency of $\sim0.16$.

\end{abstract}


\end{frontmatter}



\section{Introduction}

Dynamo theory seeks to explain the processes that generate magnetic fields in stars, planets, and galaxies. These  magnetic fields are sustained by currents generated by the flow of a conducting fluid \cite{roberts1992dynamo, weiss2002dynamos}, such as molten iron in the Earth's core, or plasma in the Sun and stars. The fluid velocity, $\boldsymbol{u}$, twists, stretches, and shears field lines, counteracting the Ohmic diffusion of the magnetic field, $\boldsymbol{B}$, which would otherwise cause the field to decay \cite{moffatt1978field}.

The dynamo problem is typically modelled using the induction equation 
\begin{equation}
\frac{ \partial \boldsymbol{B}}{\partial t} =\del \times \left( \boldsymbol{u} \times \boldsymbol{B} \right)  +  \frac{1}{R_m} \nabla ^2 \boldsymbol{B},
\label{eq:induction-dimensionless-1}
\end{equation}
coupled with the the momentum equation for the fluid, which determines $\boldsymbol{u}$.
The magnetic Reynolds number is defined by
\begin{equation}
R_m=\frac{UL}{\eta},
\label{eq:Rm}
\end{equation}
where $U$ is a characteristic speed, $L$ a characteristic length scale, and $\eta$ is the magnetic diffusivity \cite{roberts1967introduction}.
A solution of this coupled system is called a dynamo if the magnetic field continues to be generated as time $t \rightarrow \infty$. 
The smallest scale of magnetic features created by the evolution of the induction equation \eqref{eq:induction-dimensionless-1} is the resistive scale which is proportional to ${R_m}^{-1/2}$ \cite{moffatt1978field}, so that the spatial resolution required to capture these structures increases as ${R_m}^{1/2}$.
This leads to very high computational requirements that limit the parameter regimes that can be studied.
Simulations are currently carried out at parameter regimes far removed from those found in natural dynamos of real stars and planets \cite{davies2011scalability}.

Owing to the complexity of the full dynamo problem much attention has focused on the simpler problem of kinematic dynamos where $\boldsymbol{u}$ is prescribed. 
As the induction equation is linear in $\boldsymbol{B}$, solutions to the kinematic dynamo problem are either exponentially growing or decaying with a well defined growth-rate for a statistically stationary $\boldsymbol{u}$.
If $\boldsymbol{u}$ is steady, then the growth-rate can be found by solving an eigenvalue problem, whilst if the flow is periodic in time, the growth-rate can be found by solving a Floquet problem. 
If in addition the flow is independent of one co-ordinate, say the $z$-co-ordinate (2.5D flow), then one can seek monochromatic solutions of the form $\boldsymbol{B}(x,y,z,t)=\boldsymbol{b}(x,y,t)\e ^{\ii k_z z}$, so that $k_z$ becomes a parameter and the problem becomes 2D. 
In particular, there is much interest in the behaviour of dynamos at different $R_m$. 
Low $R_m$ systems are dominated by diffusion, while high $R_m$ systems are dominated by advection. 
Laboratory and engineering type flows tend to have $R_m < 1$~\cite{knaepen2008magnetohydrodynamic}, flows in the Earth's core are characterised by $R_m \sim 10^2-10^3$ \cite{kono2002recent}, whilst flows in the Sun have $R_m \sim 10^6 - 10^{10}$~\cite{Ossendrijver2003}. 

Behaviour of dynamos at large $R_m$ is of particular interest astrophysically. If the growth-rate stays bounded away from zero as $R_m \rightarrow \infty$  then a dynamo is called \textit{fast}. 
Otherwise it is called \textit{slow}. 
It is a necessary (but not sufficient) condition that a fast dynamo have chaotic Lagrangian particle paths. 
In 2.5D flows, this can only be the case for unsteady flows. 
In fully 3D flows, the paths tend to be chaotic even without time dependence. 
No fast dynamos have yet been proven mathematically, but some have been found to act as fast dynamos in numerical simulations.

A particular choice of 2.5D cellular steady flow was considered by Roberts \cite{roberts1972dynamo}.  The Galloway-Proctor circularly polarised (CP) \cite{galloway1992numerical} flow is an extension of this flow to include time dependence. While the Roberts flow must act as a slow dynamo, as it is steady, the Galloway-Proctor dynamo is thought to be fast. The Roberts flow dynamo has been used to investigate the experimental dynamo at Karlsruhe \cite{plunian2009harmonic}, whilst the Galloway-Proctor flow was used to investigate the formation of large scale magnetic fields at high $R_m$ \cite{tobias2013shear}.

High accuracy solutions are needed for dynamo simulations and so the majority of studies use spectral methods to discretise in space~\cite{jones2008course}, expanding in Fourier series, spherical harmonics, or Chebyshev polynomials as best fits the domain, to make use of their spectral convergence properties.
Because of the high resolution requirements, very long compute times are found for the full dynamo system, even when simulations are run on large numbers of cores in high performance computing (HPC) facilities.
Matsui et al.~\cite{matsui2016performance}, for example, tested scaling  on up to 16,384 cores. 
Schaeffer et al.~\cite{schaeffer2017turbulent} ran simulations that required over 10 million cpu hours.
Transforming between spectral and spatial coordinates, usually via Fast Fourier Transform (FFT), acts as a limiting factor on the parallel scalability of pseudospectral codes~\cite{mininni2011hybrid}.
Owing to the global communications requirements of spectral methods, simulations are currently unable to scale to the huge number of processors available in modern HPC facilities.
Further ways to parallelize computations are therefore required to investigate more realistic parameter regimes.

Parallel in time methods can increase  scalability of computer simulations beyond saturation of spatial parallelisation~\cite{croce2014parallel} and have been investigated for over 50 years~\cite{gander201550}. 
The most widely studied parallel-in-time algorithm is Parareal~\cite{lions2001resolution}, which was introduced in 2001 and has spurred a renewed interest in the subject. 
Further parallel in time methods have been proposed recently, like PFASST \cite{minion2010hybrid}, PITA \cite{CortialFarhat2009}, and Paraexp \cite{gander2013paraexp}. 
The Parareal algorithm has been utilised in fluid flow problems \cite{fischer2005parareal,croce2014parallel}, plasma physics \cite{samaddar2017temporal}, financial simulations \cite{bal2002parareal}, and in planetary mantle simulations \cite{samuel2012time}. 
The method has been extensively analysed mathematically by Gander and Vandewalle~\cite{gander2007analysis}, and is thought to perform badly for purely hyperbolic and highly advective systems \cite{steiner2015convergence}, as it primarily corrects amplitude defects, rather than phase or frequency defects~\cite{Ruprecht2018}. 
While some recent works demonstrate efficient application of Parareal to hyperbolic PDEs~\cite{dai2013stable}, the applicability of the method for dynamo simulations has not yet been studied.

The present paper studies the ability of Parareal to speed up kinematic dynamo simulations. 
It presents the first demonstration that parallel-in-time methods can deliver speedup for the induction equation beyond the saturation point of spatial parallelization.
We introduce an implementation of the Parareal algorithm in the open source spectral solver Dedalus \cite{burns2016dedalus}.
and investigate its performance for the stationary Roberts~\cite{roberts1972dynamo} as well as the time-dependent Galloway-Proctor~\cite{galloway1992numerical} flow. 
Although these flows are relatively simple, they generate complex dynamics in the magnetic field and are good test problems to demonstrate Parareal's efficiency for dynamo simulations. 

\section{The kinematic dynamo problem}
\label{sec:problem_formulation}

In this work, the kinematic dynamo is studied for a subset of the ABC (Arnold, Beltrami and Childress) class of flows
\begin{equation}
    \label{eq:abc_flow}
    \boldsymbol{u} = \left(C \sin(z)+B \cos(y), A \sin(x) + C \cos(z), B \sin(y) + A \cos(x) \right),
\end{equation}
\cite{galloway1984numerical}, with one of $A$, $B$, or $C$ equal to 0, making the flow 2.5D.

\subsection{Roberts Flow}

The Roberts flow is found from the ABC flow by setting $A=B=1$, $C=0$, giving $\boldsymbol{u}=\left(\cos(y),\sin(x),\sin(y)+\cos(x)\right)$. Noting that $\del \cdot \boldsymbol{B} = \del \cdot \boldsymbol{u}=0$, equation \eqref{eq:induction-dimensionless-1} becomes
\begin{subequations}
\label{eq:induction_roberts}
\begin{align}
\begin{split}
\partial_t b_x(x,y,t)= -b_y \sin(y) -\cos(y)\partial_x b_x - \sin(x)\partial_y b_x \\ - (\sin(y)+\cos(x))\ii k_z b_x+\frac{1}{R_m}\left(\nabla^2-k_z^2 \right)b_x,
\label{eq:roberts-dynamo_x}
\end{split}
\end{align}
\begin{align}
\begin{split}
\partial_t b_y(x,y,t)=b_x \cos(x)  -\cos(y)\partial_x b_y- \sin(x) \partial_y b_y \\ -   (\sin(y)+\cos(x))\ii k_z b_y+\frac{1}{R_m}\left(\nabla^2-k_z^2 \right)b_y,
\label{eq:roberts-dynamo_y}
\end{split}
\end{align}
\begin{align}
\begin{split}
    \frac{\partial b_x}{\partial x} + \frac{\partial b_y}{\partial y} + \ii k_z b_z = 0.
\end{split}
\end{align}
\end{subequations}
This flow is periodic in $x$ and $y$, and equations \eqref{eq:induction_roberts} are solved in a $2\pi$ square plane with periodic boundary conditions.

\subsection{Galloway-Proctor Flow}

The Galloway-Proctor circularly polarised (CP) flow adds time dependence to a Roberts like flow
\begin{align}
    \begin{split}
    \boldsymbol{u}= \left( C \sin(z+\sin \omega t) + B \cos(y+ \cos \omega t), \right. \\ \left. C \cos (z+ \sin \omega t ), B \sin (y + \cos \omega t)  \right) ,
    \end{split}
\end{align}
with $\omega=1$ and $C=B=\sqrt{3/2}$, $A=0$. 
We look for solutions of the form $\boldsymbol{B}=\boldsymbol{b}(y,z,t) \e^{\ii k_x x}$. When put into \eqref{eq:induction-dimensionless-1}, we have
\begin{subequations}
\begin{equation}
\partial_t b_y(y,z,t)=  b_z\partial_z v -u \ii k_x b_y  -v\partial_y b_y - w\partial_z b_y+\frac{1}{R_m}\left(\nabla^2 -{k_z}^2 \right)b_y,
\label{eq:gp-dynamo_y}
\end{equation}
\begin{equation}
\partial_t b_z(y,z,t)=b_y\partial_y w  -u \ii k_x b_z  - v\partial_y b_z - w\partial_z b_z+\frac{1}{R_m}\left(\nabla^2 - {k_x}^2 \right)b_z,
\label{eq:gp-dynamo_z}
\end{equation}
\begin{equation}
   \ii k_x b_x+ \frac{\partial b_y}{\partial y} + \frac{\partial b_z}{\partial z}  = 0.
\end{equation}
\end{subequations}

\subsection{Morphology of the Roberts and Galloway-Proctor fields}

\begin{figure}
\centering

\begin{subfigure}{0.5\linewidth}
    \includegraphics[width=\textwidth]{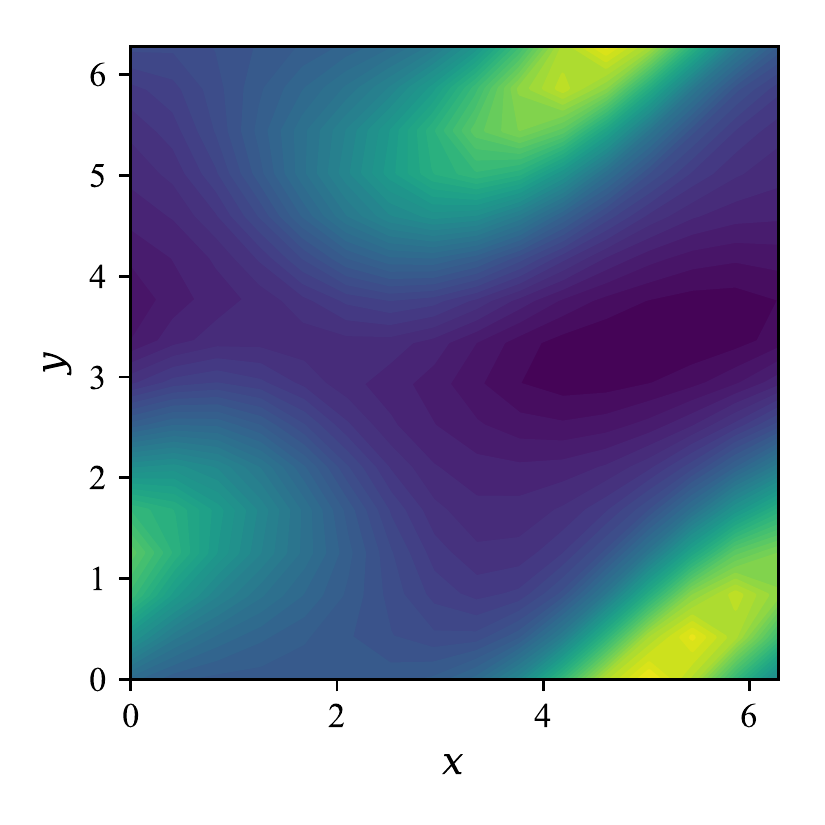}
    \caption{Roberts flow, $R_m=4$}
    \label{fig:contour_ro_4}
\end{subfigure}%
\begin{subfigure}{0.5\linewidth}
    \includegraphics[width=\textwidth]{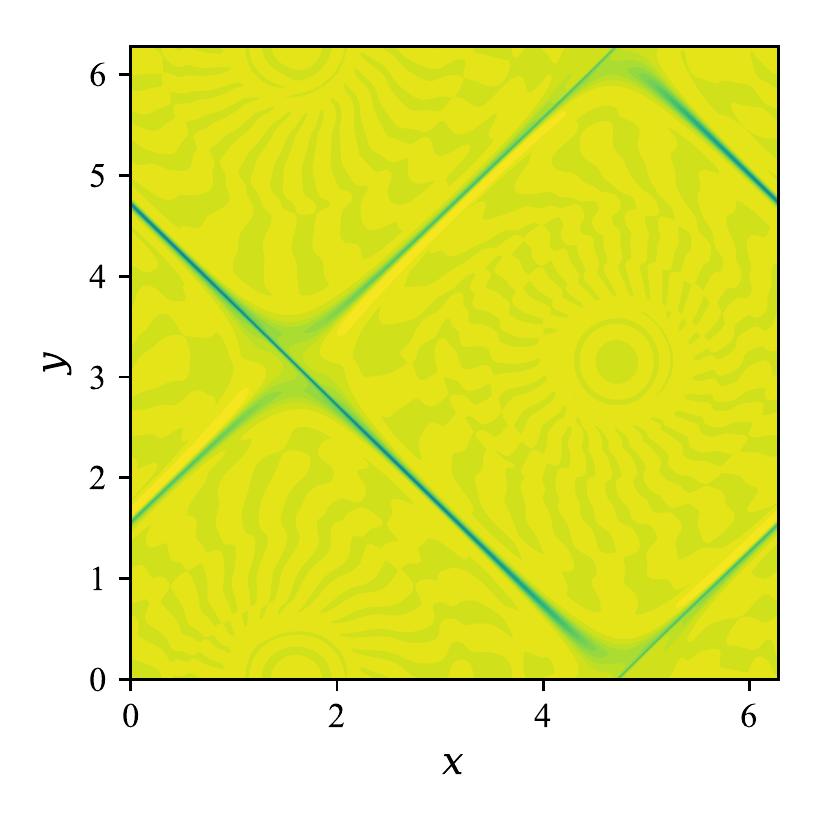}
    \caption{Roberts flow, $R_m=4096$}
    \label{fig:contour_ro_4096}
\end{subfigure}
\begin{subfigure}{0.5\linewidth}
    \includegraphics[width=\textwidth]{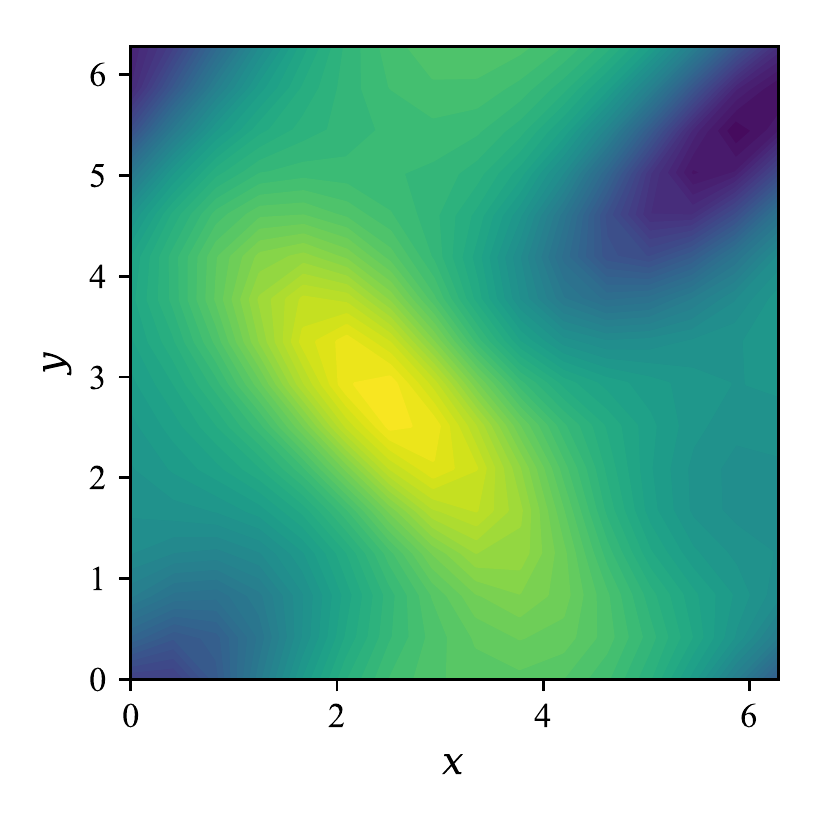}
    \caption{Galloway-Proctor flow, $R_m=3$}
    \label{fig:contour_gp_3}
\end{subfigure}%
\begin{subfigure}{0.5\linewidth}
    \includegraphics[width=\textwidth]{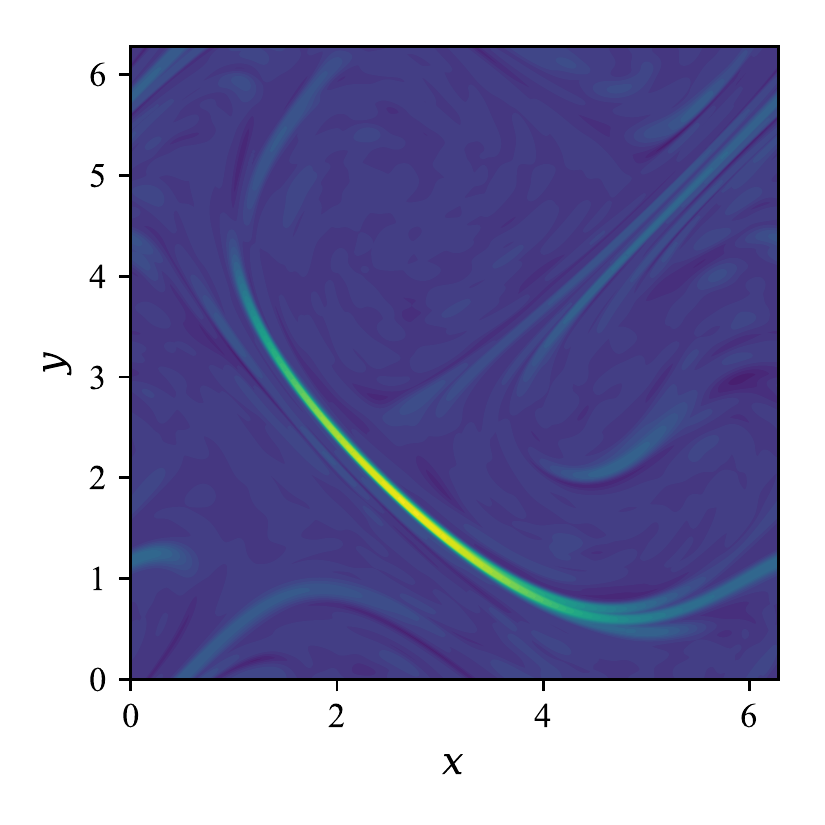}
    \caption{Galloway-Proctor flow, $R_m=3000$}
    \label{fig:contour_gp_3000}
\end{subfigure}%

\caption{Contour plots of the y-component of the magnetic fields for the labelled flows at time $T=50$. The left hand side shows the low $R_m$ field whilst the right hand plots show the high $R_m$ fields. Much finer structures are apparent in the high $R_m$ cases, due to the ${R_m}^{-1/2}$ scaling of the spatial structures. The Galloway Proctor field shows more spatial variability, due to the time dependence of the flow. The Roberts field stays effectively fixed in space, only varying in magnitude over the course of the simulation, whilst the morphology and magnitude of the Galloway Proctor field changes over time.}
\label{fig:contours}
\end{figure}

Figure \ref{fig:contours} shows the contours of the y-component of the magnetic field for $R_m=3$ and $3000$ in the Galloway-Proctor flow, and $R_m=4$ and $4096$ in the Roberts flow. 
Larger scale and more diffuse structures are present in the $R_m=3$ and $R_m=4$ simulations. 
Finer structures emerge in the $R_m=3000$ and $R_m=4096$ cases, showing the effect of the ${R_m}^{-1/2}$ scaling on the smallest structures. 
The Galloway-Proctor magnetic field morphology changes over time, with the periodic change in the velocity field, which leads to the more complicated pattern shown in Figure~\ref{fig:contour_gp_3000}. For the Roberts flow, the magnetic field has a constant morphology and simply grows exponentially in magnitude.

\section{Parareal Algorithm}\label{sec:parareal}
We give only a brief description of the Parareal algorithm, for a more detailed description see for example the works by Lions et al. or Gander and Vandewalle~\cite{lions2001resolution,gander2007analysis}. 
Parareal is an iterative method for solving initial value problems (IVPs) of the form 
\begin{equation}
    \frac{\partial U}{\partial t} = f\left(U(t),t\right), ~~ U(0)=U_0, ~~ 0\leq t \leq T
\end{equation}
where the right hand side $f$ typically comes from spatially discretizing a partial differential equation.
Parareal uses two different time stepping methods, the coarse method $\mathcal{G}$ and the fine method $\mathcal{F}$. 
The time domain is split up into $N_p$ time slices, defined by the time-points $0=t_0<t_1<...<t_{N_p-1}<t_{N_p}=T$, where $N_p$ is the number of processors available for time parallelisation. 
We assume that all time slice are of equal length $\Delta T=T/N_p$. 
The fine time stepping method has time step $\delta t$, while the coarse method has time step $\Delta t$.

Parareal starts with running the coarse solver from $t=0$ to $t=T$, giving an initial approximation to the solution $U$ at time $t_n$, $U^{k=0}_n$, for every $n=0,1,...,N_p$, where  $k$ denotes the Parareal iteration number, and $n$ denotes the time slice. 
Then, the approximation is refined using the iteration
\begin{equation}
U_{n+1}^{k+1} = \mathcal{G}(t_{n+1},t_n,U_n^{k+1})+\mathcal{F}(t_{n+1},t_n,U^k_n)-\mathcal{G}(t_{n+1},t_n,U_n^k).
\label{eq:parareal}
\end{equation}
Computing the fine method can be parallelized across time slices.
However, the correction from the coarse propagator has to be computed in serial, propagating the Parareal correction throughout the time domain.

\subsection{Convergence and performance}
As $k \rightarrow N_p$, Parareal converges to the solution that would have been obtained through the use of the fine method in serial over the time domain $[0,T]$~\cite{gander2007analysis}. 
Achieving speed up, however, is dependent upon converging to an acceptable tolerance ${tol}$ within a much smaller number of iterations than $N_p$. 
We test for convergence by measuring the relative two-norm defect at the last time slice between successive iterations 
\begin{equation}
  \sigma = \frac{\left\|U^{k}_{N_p}-U^{k-1}_{N_p}\right\|_2}{\left\|U^{k}_{N_p}\right\|_2}.
  \label{eq:error}
\end{equation}
We denote as $k_{con}$ the number of iterations required until $\sigma \leq {tol}$.

Speed up $s$ of Parareal can be estimated as 
\begin{equation}
    s = \left[ \left( 1+ \frac{k_{con}}{N_p} \right) \frac{R_c}{R_f} + \frac{k_{con}}{N_p} \right]^{-1},
    \label{eq:speed_up}
\end{equation}
where $R_c$ is the runtime of the coarse method over one time slice and $R_f$ is the runtime of the fine method over one time slice. 
Speed up is bounded by
\begin{equation}
    s \leq \min \left\{ \frac{N_p}{k_{con}}, \frac{R_f}{R_c} \right\},
\label{eq:speed_bound}
\end{equation}
\cite{gander2007analysis}.
The bound illustrates the trade off that needs to be optimised to gain optimal performance with Parareal. 
On the one hand, the ratio between the run times of the fine and coarse solvers should be large, so that the second bound is high.
This can be achieved by using a very coarse and cheap $\mathcal{G}$.
However, the method also needs to converge in a small number of iterations to ensure that the first bound is high. 
A less accurate coarse solver will typically require a larger number of iterations to converge. 
Finding a good compromise between these two competing factors is key to gaining high speed up with Parareal. 
Parallel efficiency is defined as 
\begin{equation}
    \epsilon = \frac{s}{N_p} = \frac{\text{parallel speed up}}{\text{number of processors}},
\end{equation}
where ideal scaling $s = N_p$ would give an efficiency of one. 
Parareal has an efficiency bound of $1/k_{con}$, which highlights the need to keep the number of iterations low. 
The need to test for convergence between two successive iterations of Parareal means there is an effective limit of $1/2$ for parallel efficiency when using Parareal.

\subsection{Coarse solvers}
A number of strategies have been proposed to design coarse solvers for Parareal. 
The most simple to implement is an increase in time step size for the coarse solver compared with the fine solver, whilst leaving the timestepping method and spatial discretization unchanged~\cite{croce2014parallel,aubanel2011scheduling}. 
A second strategy is to use an implicit time stepping method with large step size for the coarse solver, with explicit time stepping for the fine solver~\cite{blouzaetal2009}.
Further strategies to find a coarse solver include simplifying the physics so that a less complicated model can be used~\cite{baffico2002parallel,maday2003parallel,maday2007parareal}.
Another strategy to be considered is to use a reduced spatial resolution, along with a larger time step~\cite{lunet2018time}. Using this strategy, the method of interpolation from coarse to fine grids has been found to be important to Parareal convergence~\cite{ruprecht2014convergence}. 
In section~\ref{sec:coarse_solver_study} we explore different options in the context of the kinematic dynamo problem.

\section{Implementation}\label{sec:implementation}

This work was carried out using the Dedalus~\cite{burns-prep-dedalus} open source, spectral solver. 
We use a collocation-based pseudo-spectral method.
Spectral methods benefit from spectral convergence: for a sufficiently smooth solution, the error of the method is $\mathcal {O}[(1/N_s)^{N_s}]$ for a discretisation with $N_s$ collocation points. 
This compares favourably with a standard finite difference discretisation, which has error $\mathcal{O}[(1/N)^p]$ for a $p^{\text{th}}$ order method. 
Because of the periodic domain, Fourier bases are used.
Dedalus uses the FFTW library to perform fast parallel transforms between real and spectral space and parallelizes in space over $n-1$ dimensions of an $n$ dimensional domain using the mpi4py library~\cite{dalcin2005mpi}.

Dedalus offers a number of different time stepping methods up to \nth{4} order, including multi-step and Runge Kutta IMEX methods~\cite{ascher1997implicit,spalart1991spectral,wang2008variable}.  
Because multi-step methods in Parareal require restarting in every time slice and every iteration, we focus on Runge-Kutta methods.
For optimal serial performance, we rely on implicit/explicit (IMEX) Runge-Kutta methods~\cite{ascher1997implicit}.
Here, we integrate the terms $(\boldsymbol{u} \cdot \del)\boldsymbol{B}$ and $(\boldsymbol{B} \cdot \del)\boldsymbol{u}$ explicitly, whilst the diffusion term will be treated implicitly.
Below, we compare the IMEX Runge-Kutta methods RK111 (1 implicit stage, 1 explicit stage, \nth{1} order), RK222 (2 implicit stages, 2 explicit stages, \nth{2} order), and RK443 (4 implicit stages, 4 explicit stages, \nth{3} order) to determine the most efficient serial fine solver against which to compare Parareal.

Simulations are carried out in a periodic domain with side length $2\pi$.
The spatial resolution is measured in terms of the number of spectral modes in $x$ and $y$,  ($N_x$, and $N_y$ respectively) for the Roberts flow, or number of modes in $y$ and $z$, ($N_y$ and  $N_z$ respectively) for the Galloway-Proctor flow. 
Since the domain is a square, we set $N_x = N_y = N$ for the Roberts flow and $N_y=N_z=N$ for the Galloway-Proctor flow in all examples, so that the total spatial problem size is $N^2$. 
The initial conditions were set to a random perturbation of magnitude $10^{-5}$ for $B_x$ and $B_y$ in the Roberts flow, and for $B_y$ and $B_z$ in the Galloway Proctor flow.

In the kinematic approach, the simulation time needs to be long enough that the largest growing mode can be found and a steady growth rate calculated. 
In this work, this time was found to be around 10-20 turnover time periods. 
However, in a fully dynamic simulation, much longer simulation times are required to reach a steady state in both the fluid flow and the magnetic field. 
For example, Smith et al.~\cite{smith2004vortex} run simulations until $T=450$ to achieve a statistically steady flow. 
As the aim of the work is to inform future studies of dynamic simulations, we choose a longer time interval of 50.

Parareal was implemented by splitting the MPI world communicator into a space communicator and a time communicator. The space communicator was utilised by Dedalus to parallelize in space, whilst the time communicator was used to communicate between time slices. Two instances of Dedalus were created on each time slice, one with the coarse resolution ($N_C$ spectral modes in each direction) and time step $\Delta t$, and one with the fine resolution ($N_F$ spectral modes in each direction) and time step $\delta t$. 
Interpolation from coarse to fine grids, and restriction from fine to coarse grids, were carried out using the Dedalus \pyth{set_scales(ratio)} method on the field objects, which allows efficient interpolation based on Fourier re-sampling.

\subsection{Pseudo-code for the Parareal algorithm in Dedalus}

\begin{algorithm}[]
 \tcp{Initialization}
 space\_communicator=MPI.COMM\_WORLD.Split(time\_slice, key)\;
 time\_commnicator=MPI.COMM\_WORLD.Split(space\_slice, key)\;
calculate time slice size, start and end of time slice for each time slice\;
create 2 instances of dedalus solver on each time slice, using the space communicator, with same equations, but different time step sizes and resolutions\;
set initial conditions of fine solver\;
restrict initial conditions down to size of coarse solver, and set coarse solver initial conditions\;
\end{algorithm}

\begin{algorithm}[]
\tcp{Initial coarse run}
 \For{each time slice from 0 to T\_end}{
   \eIf{first time slice}{
     get initial conditions
   }{
   
   time\_communicator.Recv(coarse\_fields,source=time\_communicator.rank-1)\;
   \tcp{next step gets fine solver ready for parareal iterations}
   
   coarse\_fields.set\_scales(N\_fine/N\_coarse)\;
   fine\_fields=np.copy(coarse\_fields)
   }
  \For{ i in range(N\_time\_steps\_per\_slice\_coarse)}{ 
   coarse\_solver.step(coarse\_time\_step)}
   time\_communicator.Send(coarse\_fields,dest=time\_communicator.rank+1)\;
 }
 \end{algorithm}
 
 \begin{algorithm}[]
 
 \tcp{Beginning of Parareal iterations}
 \While{not converged}{
   \tcp{Parallel step}
   \For{ i in range(N\_time\_steps\_per\_slice\_fine)}{ 
   fine\_solver.step(fine\_time\_step)}
   
   \tcp{Serial Step}
   \For{every time slice}{
     \eIf{first time slice}{
       get initial conditions
     }{
       time\_communicator.Recv(fine\_fields,source=time\_communicator.rank-1)\;
     }
     \tcp{Parareal correction}
   fine\_fields.set\_scales(N\_coarse/N\_fine)\;
   coarse\_fields=np.copy(fine\_fields)\;
   \For{ i in range(N\_time\_steps\_per\_slice\_coarse)}{ 
   coarse\_solver.step(coarse\_time\_step)}
   carry out correction (new\_coarse - old\_coarse + new\_fine)\;
   fine\_fields = new\_coarse\_result - old\_coarse\_result + fine\_fields\;
   \If{not last time slice}{
   time\_communicator.Send(fine\_fields,dest=time\_communicator.rank+1)\;
   }
   save corrected solution\;
   set time to t\_start for each slice (coarse and fine)\;
   }
   \If{final time slice}{
     check for convergence with previous solution\;
   }
 } 
 \end{algorithm}
 \begin{algorithm}[]

 \tcp{Saving state} 
\SetKwFunction{FMain}{Save\_state} 
\SetKwProg{Pn}{Function}{:}{\KwRet}
  \Pn{\FMain{}}{
   analysis=fine\_solver.evaluator.add\_file\_handler(file\_name,iter=1)\;
   analysis.set\_num=time\_communicator.rank
   analysis.add\_system(fine\_solver.state,layout='g')\;
   fine\_solver.step()\;
   analysis.iter=np.inf\;
        
  } 
\end{algorithm}

\FloatBarrier

\section{Results}\label{sec:results}
To ensure that reported speedups are a like-to-like comparison and meaningful, we need to make sure that the solution provided by Parareal is of the same accuracy as the fine solver used serially.
Given the need for highly accurate results in dynamo simulations, we aim for an accuracy of $10^{-5}$.
Furthermore, we need to compare Parareal against an efficient fine solver -- achieving speedup with Parareal when a much more efficient alternative to the fine solver exists would not provide convincing evidence of Parareal's usefulness.
Therefore, we first find the optimal fine solver in Dedalus to deliver the required accuracy and then parallelise it with Parareal.

\subsection{Fixing the fine solver}\label{sec:accuracy}

\begin{figure}[th]
\centering
\begin{subfigure}{0.5\textwidth}
\includegraphics[width=\textwidth]{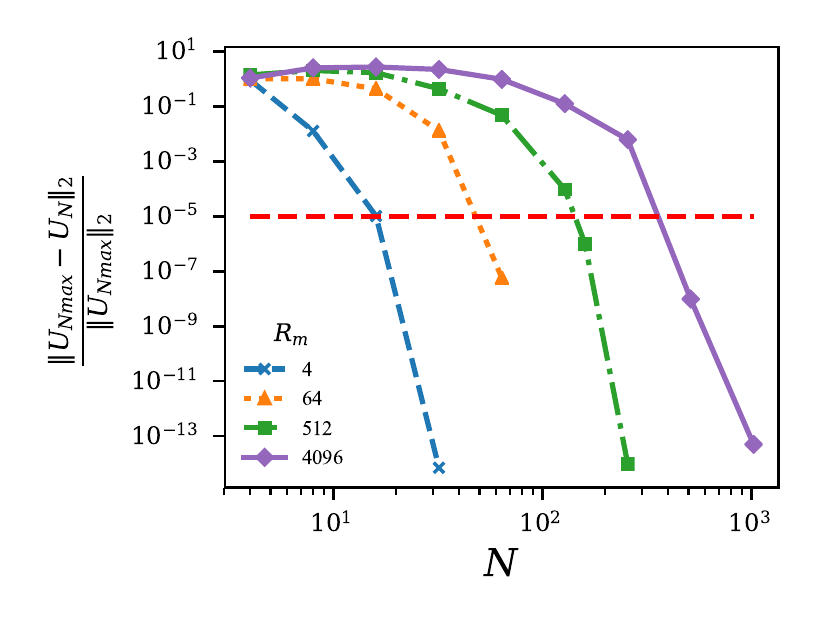}

\caption{Roberts Flow}
\end{subfigure}%
\begin{subfigure}{0.5\textwidth}
\includegraphics[width=\textwidth]{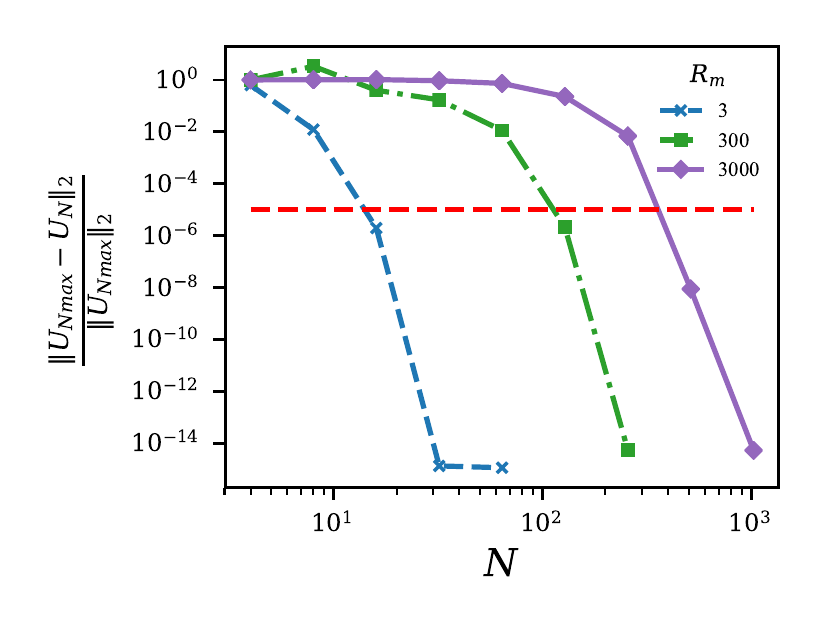}

\caption{Galloway Proctor Flow}
\end{subfigure}
\caption{Graphs showing spatial convergence of the solvers for the two flows investigated. Spectral order of convergence is observed, with the decrease in error accelerating as the number of spectral modes, $N$, is increased. The line at $10^{-5}$ shows the required level of accuracy in the solution. The number of modes required for each $Rm$ follows the predicted $Rm^{1/2}$ scaling.}
\label{fig:spatial_convergence}
\end{figure}

First, we need to fix the required spatial resolution for $\mathcal{F}$.
Convergence in space was tested by running simulations at double the previous spatial resolution until the normalised $L^2$ difference between two solutions was $ \sim 10^{-15}$.
At this point, the error from the spatial discretisation is of the order of machine precision. 
We denote $U_N$ as the solution vector containing $b_x$ and $b_y$ in real space with resolution $N$. 
The smaller solution was interpolated onto the same grid size as the fine solution using spectral interpolation so that the difference could be computed.
The most highly resolved solution, $U_{Nmax}$, was then used as a reference solution to compute relative error of each $U_N$. 
The results are shown in Figure~\ref{fig:spatial_convergence}, confirming the expected spectral convergence behaviour. 
For each $R_m$, we set $N_F$ to the smallest value that gives a solution with error smaller than $10^{-5}$ (indicated by the dashed red line).
Because higher magnetic Reynolds numbers produce smaller scale features, they require better spatial resolution to match our error tolerance.

\begin{figure}[ht]
\centering
\includegraphics[]{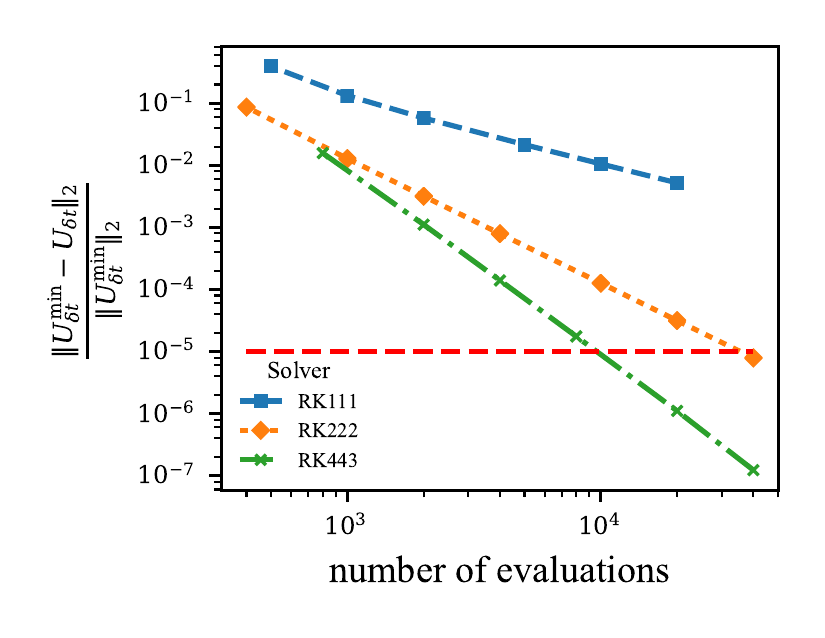}
\caption{Work required for different time stepping methods to obtain solutions of a certain accuracy, measured as the relative two-norm of the each solution $U_{\delta t}$ and the solution obtained using the smallest timestep with the RK443 stepper ($U_{\delta t}^{\text{min}}$). The number of evaluations required to compute from $T=0$ to $T=1$ are shown. This number depends on $\delta t$ and the number of stages in each time-stepping method. For any degree of accuracy better than $10^{-2}$, the RK443 time-stepper requires fewer evaluations than either RK111 or RK222. Results shown are for the Galloway-Proctor flow with $R_m=300$, $N_F=256$.}
\label{fig:dt_convergence_solvers}
\end{figure}

Next, we fix the time stepping method and time step.
Creating a reference solution with a temporal error of the order of machine precision proved to be unfeasible, due to computational constraints, especially in the higher $R_m$ cases. 
Therefore, a result with error lower than $10^{-7}$ was used as a reference solution for setting $\delta t$. 
This is two orders of magnitude smaller than the desired result of $10^{-5}$ and should provide an accurate estimation of the error due to time-stepping.
A comparison of the Runge-Kutta time steppers available in Dedalus is shown in Figure~\ref{fig:dt_convergence_solvers}. 
Because RK443 reaches the requested tolerance of $10^{-5}$ with the smallest number of evaluations of the right hand side function, it is the most efficient choice. Similar results were found for other magnetic Reynolds numbers.
We therefore use RK443 for the fine method $\mathcal{F}$ throughout this work.

\begin{figure}[t]
\centering
\begin{subfigure}{0.5\linewidth}
\includegraphics[width=\textwidth]{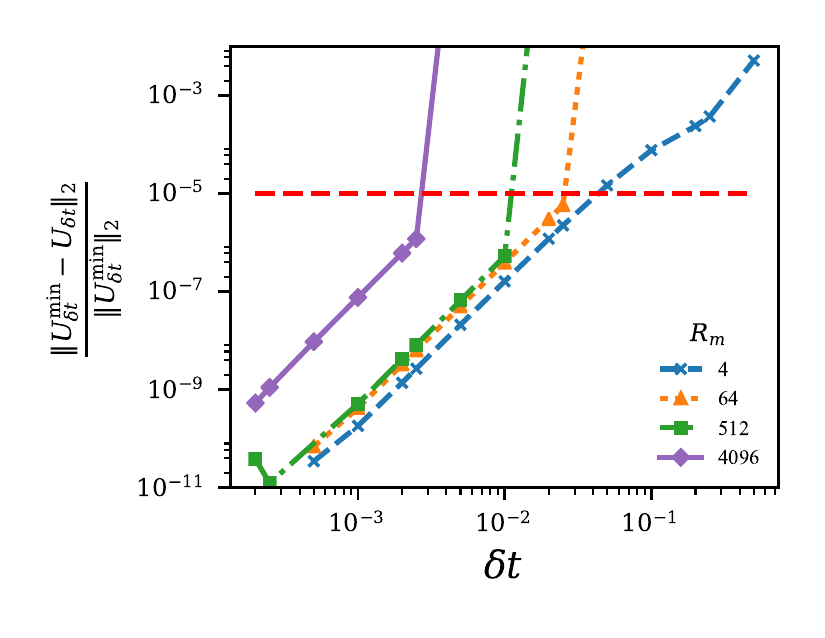}
\captionsetup{width=0.75\linewidth}
\caption{Roberts Flow}
\label{fig:dt_converge_roberts}
\end{subfigure}%
\begin{subfigure}{0.5\linewidth}
\includegraphics[width=\textwidth]{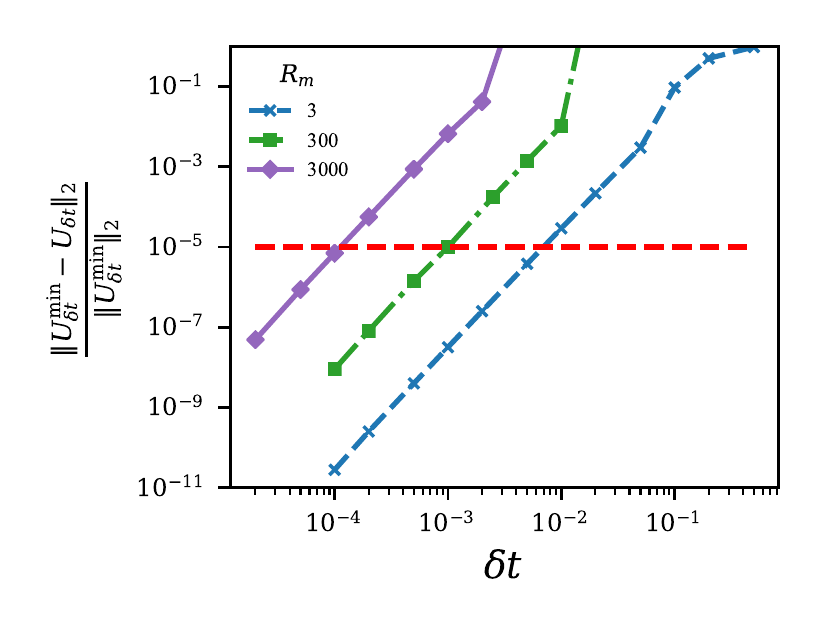}
\captionsetup{width=0.75\linewidth}
\caption{Galloway Proctor Flow}
\label{fig:dt_converge_galloway-proctor}
\end{subfigure}
\caption{Convergence with respect to time step size for the different flows and $R_m$ simulated. The Roberts flow, which is independent of time, shows high accuracy for the highest stable time step for all simulations except $R_m=4$. The Galloway Proctor flow has a larger error for the same time step size. This is believed to be because of the incorporation of time on the right hand side of the equations. Galloway-Proctor flows therefore require smaller time step sizes to reach the desired accuracy. Where the error goes past the top of the figure, the solver has diverged and is unstable for this time step size.}
\label{fig:dt_convergence}
\end{figure}

Figure~\ref{fig:dt_convergence} shows the dependence of the solution error on time step size for a range of $R_m$ for the Roberts and Galloway-Proctor flows. 
All simulations were carried out with the optimal $N_F$ found from the spatial resolution study. 
We can see that smaller step sizes are required to meet a given level of accuracy for the Galloway-Proctor flow than for the Roberts flow. 
This is likely due to the Galloway-Proctor flow depending explicitly on time. 
In the case of the Roberts flow, a $\delta t$ small enough to satisfy the stability requirements for a given $N_F$ is sufficient to also satisfy the accuracy requirement of $10^{-5}$, except for the most simple case of $R_m=4$.
This affects the performance of Parareal through the ratio of computational run times $R_f/R_c$ because we have to use essentially the same time step for both the coarse and fine method.
In contrast, for the Galloway-Proctor simulations, satisfaction of the stability requirement did not guarantee accuracy within the required tolerance, and a reduced $\delta t$ must be used for the fine solver, leading to a better coarse-to-fine computation time ratio.

\FloatBarrier

\subsection{Fixing the coarse solver}
\label{sec:coarse_solver_study}
The $N_F$ and $\delta t$ determined in Section~\ref{sec:accuracy} are used in the fine solver. 
We now discuss the different possibilities available for choosing a coarse solver. Using the same spatial resolution with  $\Delta t > \delta t$ was not suitable for the Roberts flow, as $\delta t$ was the largest stable time step for a given resolution. 
It was also unsuitable for the Galloway-Proctor simulations, as the ratio of $\Delta t / \delta t$ was not large enough to give meaningful speedup. 
Use of a fully implicit coarse solver was rejected, as the spectral spatial discretisation meant that a dense matrix solve would be required at each time step. 
This large increase in computational complexity would reduce the difference in computations required between the fine and coarse solvers, leading to smaller speed ups. 
There was little scope to attempt to use reduced physics in this study, as we are already considering the simplest form of dynamo problem. 
However, this strategy may be useful in further work on a non-linear dynamo. 
Coarsening in both space and time was found to be the most promising strategy.
As time step stability is linked to spatial resolution, reducing spatial resolution allows a larger time step to be taken, even where the fine solver is at the largest stable time step. 
This means that $N_C<N_F$ and $\Delta t > \delta t$, opening up the possibility of a large difference in computational complexity between the coarse and fine solvers.
However, too aggressive coarsening will lead to a very inaccurate coarse solver and slow convergence.
The most efficient amount of spatial coarsening was studied by carrying out Parareal simulations with a wide range of coarse method spatial resolution.

\begin{figure}[ht]
\centering
\begin{subfigure}{0.5\linewidth}
\includegraphics[width=\textwidth]{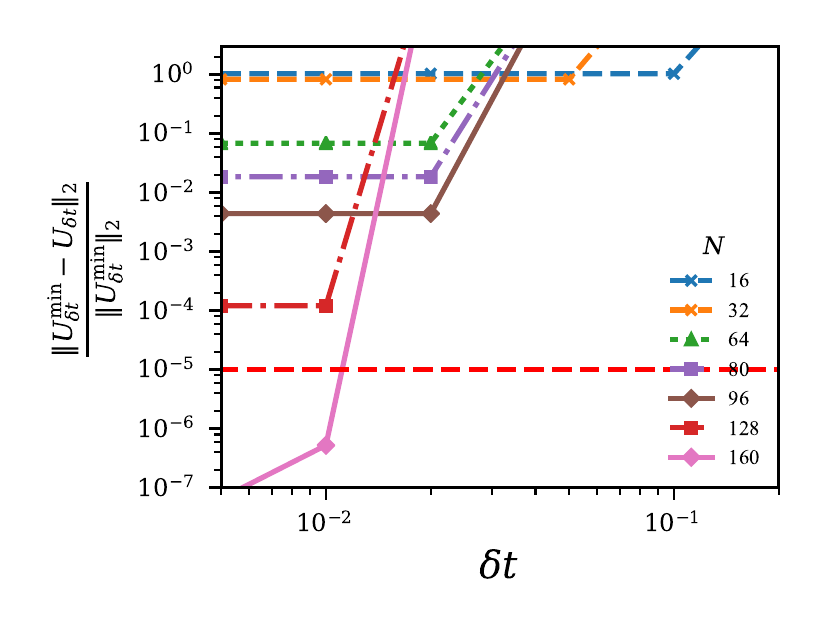}
\caption{Roberts Flow, $R_m=512$}
\label{fig:coarse_dt_stable_ro}
\end{subfigure}%
\begin{subfigure}{0.5\linewidth}
\includegraphics[width=\textwidth]{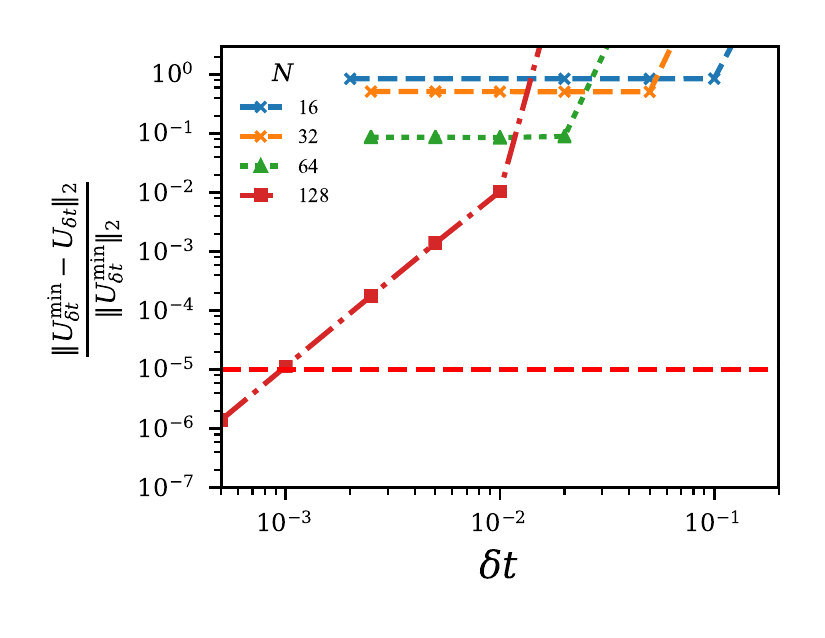}
\caption{Galloway-Proctor Flow, $R_m=300$}
\label{fig:coarse_dt_stable_gp}
\end{subfigure}
\caption{Error vs. time step size for a range of spatial resolutions ($N$) for the Roberts and Galloway-Proctor flows. Accuracy is constrained by spatial resolution, until the finest resolution is reached in each case. As the resolution reduces, the largest stable time step increases as expected. Error increases above $10^1$ (off the top of the graph) indicate that the method has become unstable at that time step. Accuracy for a given resolution/ time step size is higher for the Roberts flow than the Galloway-Proctor flow. }
\label{fig:coarse_dt_stable}
\end{figure}

\begin{figure}[ht]
\centering
\begin{subfigure}{0.5\linewidth}
\includegraphics[width=\textwidth]{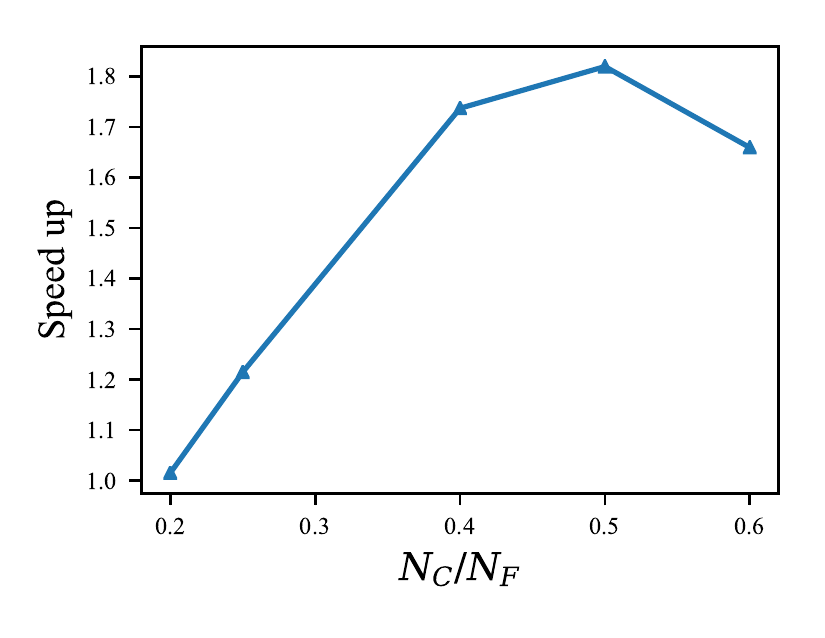}
\captionsetup{width=0.75\linewidth}
\caption{Speed up}
\end{subfigure}%
\begin{subfigure}{0.5\linewidth}
\includegraphics[width=\textwidth]{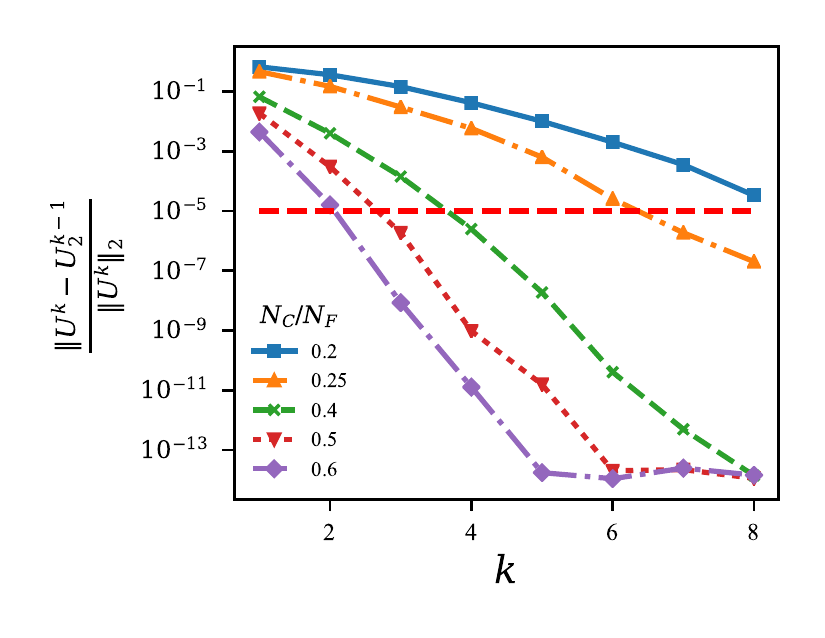}
\captionsetup{width=0.75\linewidth}
\caption{Convergence}
\end{subfigure}
\caption{(a) Speed up vs $N_C$ for Roberts flow, with $R_m=512$, $N_F=160$, and $\delta t = 10^{-2}$. Long run times are found for very low coarse resolutions as the estimated solution is not accurate enough to allow quick convergence. As $N_C$ increases, the run time reduces due to the reduced number of Parareal iterations required to converge. The best performing coarse solver has a resolution of $0.5N_F$. Further increasing the resolution of the coarse solver increases the complexity of the coarse solver to a level close to that of the fine solver, reducing any speed up possible. (b) Graph showing how defect to previous solution changes with number of Parareal iterations for different resolutions of the coarse solver. Very low resolutions results in Parareal taking many iterations to converge, reducing opportunity for speed up. High resolutions show quicker convergence.}
\label{fig:parareal_speed_up_diff_coarse}
\end{figure}

Simulations were carried out for the Roberts flow with $R_m = 512$. 
This is moderately high, whilst allowing relatively modest compute resources to be used. 
The fine solver parameters were fixed, with $N_F=160$, and $\delta t=10^{-2}$, while the coarse step $\Delta t$ was set to the highest stable step for the given $N_C$.
This was found by estimating the error at different time steps for each resolution, as shown in Figure~\ref{fig:coarse_dt_stable_ro}. A similar study was carried out for the Galloway-Proctor flow (Figure~\ref{fig:coarse_dt_stable_gp}).
The number of Parareal time slices $N_P$ was fixed at 10. 
Figure~\ref{fig:parareal_speed_up_diff_coarse} shows that the peak speed up is acquired when $N_C = 0.5 N_F$. 
When $N_C<0.5N_F$, the speed up is reduced by the extra number of Parareal iterations required to converge, and when $N_C>0.5N_F$, the difference in computational complexity between the coarse and fine solvers is insufficient.

\subsection{Scaling Results}

\begin{table}[]
\centering
\caption{Parameters for simulations. $R_m$: magnetic Reynolds number, $k_z$: wave number in z co-ordinate, $k_x$: wave number in x co-ordinate, $N_F$: number of modes in fine propagator, $N_C$: number of modes in coarse propagator, $\delta t$: time step for fine propagator, $\Delta t$: time step for coarse propagator, Growth rate indicates growth rate of the magnetic field.}
\begin{tabular}{|l|l|ll|l|l|l|l|l|}
\hline
Flow & $R_m$ & \textbf{$k_z$} & \textbf{$k_x$} & \textbf{$N_F$} & \textbf{$N_C$} & \textbf{$\delta t$} & \textbf{$\Delta t$} & Growth \\ & & & & & & & & Rate \\ \hline
Roberts & 512 & 2.87 &  & 160 & 80 & $10^{-2}$ & $2 \times 10^{-2}$ & 0.11 \\
 & 4096 & 7.5 &  & 512 & 256 & $2.5 \times 10^{-3}$ & $5 \times 10^{-3}$ & 0.097 \\ \hline
Galloway- & 3 &  & 0.57 & 16 & 8 & $5 \times 10^{-3}$ & $10^{-1}$ & 0.15 \\
Proctor & 300 &  & 0.57 & 128 & 64 & $10^{-3}$ & $2 \times 10^{-2}$ & 0.3 \\
 & 3000 &  & 0.57 & 512 & 256 & $10^{-4}$ & $5 \times 10^{-3}$ & 0.3 \\ \hline
\end{tabular}
\label{tab:pars}
\end{table}

Scaling tests were carried out for both the Roberts flow and the Galloway-Proctor flow. 
Simulations of the Roberts flow were carried out on the ARC 3 HPC facility at the University of Leeds, made up of Intel Xeon E5-2650v4 (Broadwell) CPU's, with a total of 6,048 cores. Simulations of the Galloway-Proctor flow were carried out on the ARCHER HPC facility, made up of Intel Xeon E5-2697v2 (Ivy Bridge) CPU's, with a total of 109,056 cores.

A range of $R_m$ were simulated, to see the effect on Parareal performance (see Table~\ref{tab:pars}). 
Scaling performance was compared with pure spatial scaling of the Dedalus solver. 
Fully parallel in space and in time simulations were also carried out in order to show how Parareal can increase scalability beyond the saturation of spatial scaling. 
Validation was carried out by comparing computed growth rates with those found in the literature. 
Growth rates for the Roberts dynamo were found to be consistent with those found by Plunian and Radler~\cite{plunian2009harmonic}, which were reported to 2 significant digits, with the peak growth rate for each $R_m$ occurring at the same $k_z$ wave number reported. 
Growth rates for the Galloway-Proctor simulations were consistent with those found by Charbonneau et al.~\cite{charbonneau2012solar}, with the peak growth rate occurring at a $k_x$ of 0.57. 
The Galloway-Proctor flow had the correct behaviour in terms of growth rate for large $R_m$, with the growth rate staying positive, showing the expected fast-dynamo behaviour.

\subsection{Roberts Flow}
\begin{figure}[ht]
\centering
\begin{subfigure}{0.5\linewidth}
\includegraphics[width=\textwidth]{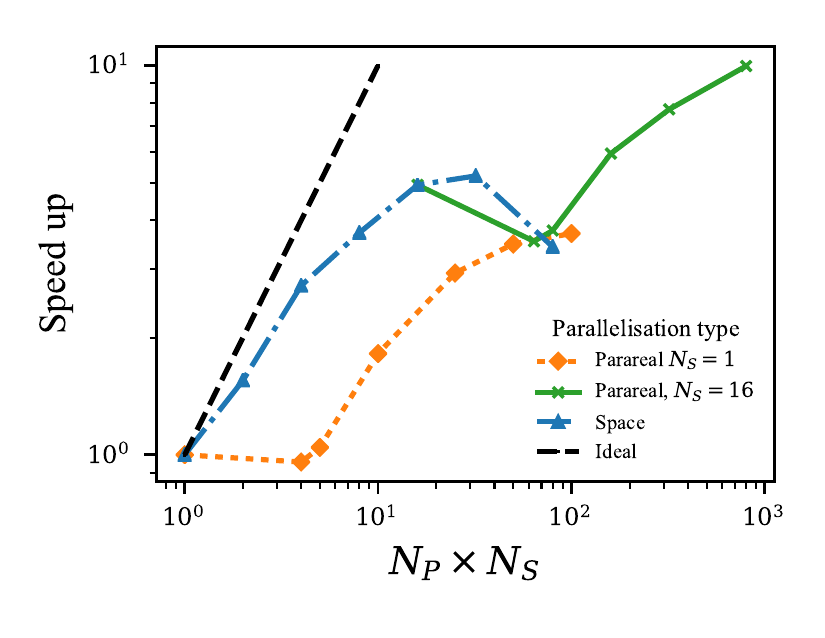}
\captionsetup{width=0.75\linewidth}
\caption{Speed up, $R_m=512$}
\end{subfigure}%
\begin{subfigure}{0.5\linewidth}
\includegraphics[width=\textwidth]{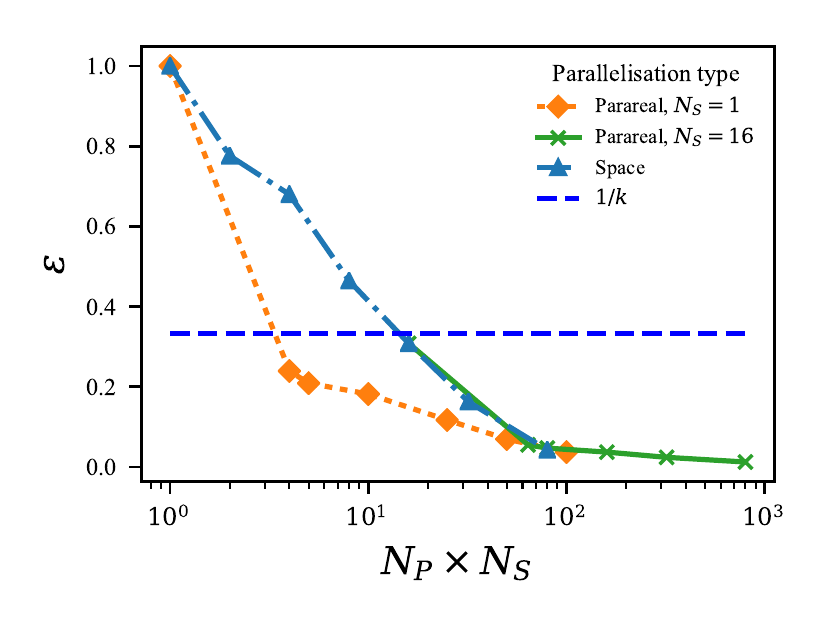}
\captionsetup{width=0.75\linewidth}
\caption{Efficiency, $R_m=512$}
\end{subfigure}
\begin{subfigure}{0.5\linewidth}
\includegraphics[width=\textwidth]{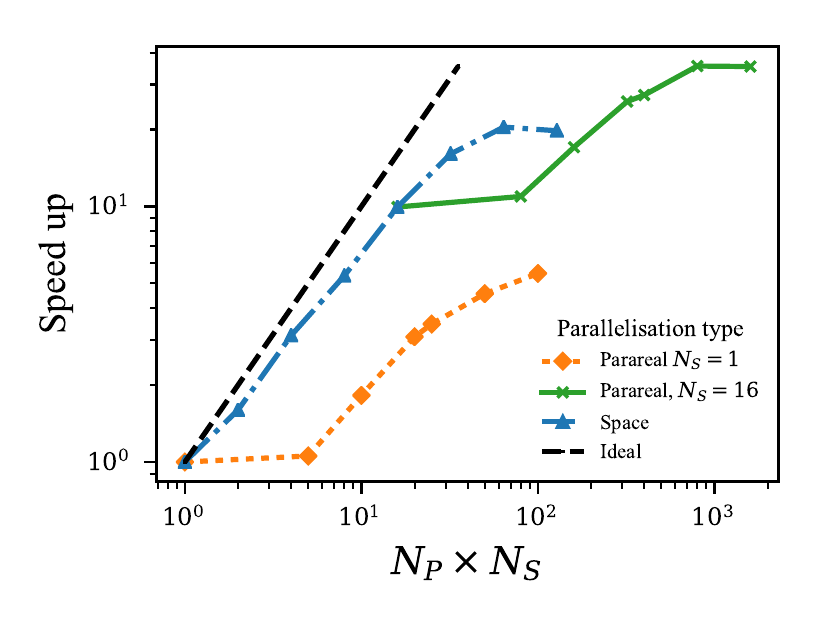}
\captionsetup{width=0.75\linewidth}
\caption{Speed up, $R_m=4096$}
\end{subfigure}%
\begin{subfigure}{0.5\linewidth}
\includegraphics[width=\textwidth]{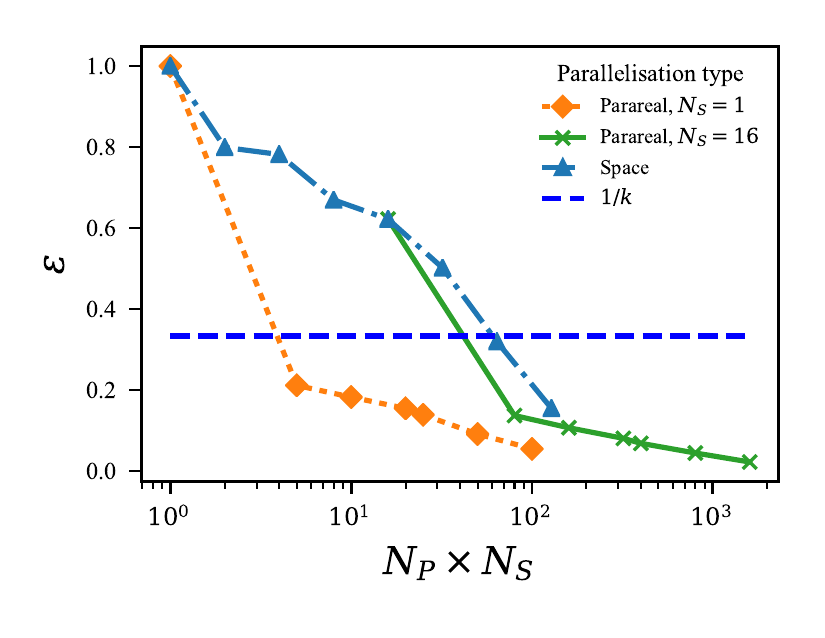}
\captionsetup{width=0.75\linewidth}
\caption{Efficiency, $R_m=4096$}
\end{subfigure}
\caption{Speed up (a), (c) and parallel efficiency (b), (d) of the Parareal method compared with spatial parallelisation for simulations of Roberts flow with $Rm$=512 (a), (b) and $R_m$=4096 (c), (d). Total number of processors is calculated as number of processors in space ($N_S$) multiplied by number of processors for Parareal ($N_P$). Speed up and efficiency are both poor for low numbers of processors for Parareal, but as the parallelisation in space saturates, further gains can be made from Parareal, although they are small. Parareal does not offer any gain over parallelisation in space until parallel efficiency is less than 0.1, and does not come close to the theoretical maximum of $1/k$, where $k$ is number of Parareal iterations.}
\label{fig:rob_speed_up_efficiency}
\end{figure}

Scaling results for the Roberts flow are shown in Figure~\ref{fig:rob_speed_up_efficiency} for $R_m = 512$ (upper figures) and $R_m = 4096$ (lower figures).
For both cases, both space parallel scaling and efficiency are superior to Parareal in the beginning.
As expected, spatial scaling is better for the $R_m = 4096$ case with higher spatial resolution, due to higher workload per processor.
While Parareal alone is not competitive, in both cases a combined space-time parallelization generates slightly more speedup than a pure spatial parallelization.
The theoretical maximum efficiency for Parareal is $1/3$, indicated by a horizontal dashed line, due to the simulation requiring three iterations to converge. 
However, because of the relatively expensive coarse solver, Parareal's observed efficiency is mostly substantially lower.
As the efficiency of the combined space-time parallelisation is the product of the parallel in space efficiency and the parallel in time efficiency, it is low for high numbers of processors because of the low efficiency of Parareal for the Roberts flow.
Despite the larger overall speedup, with efficiencies below 0.1, space-time parallelization using Parareal may not be particularly attractive .

\subsection{Galloway Proctor Flow}
\begin{figure}[ht]
\centering
\begin{subfigure}{0.49\linewidth}
\includegraphics[width=\textwidth]{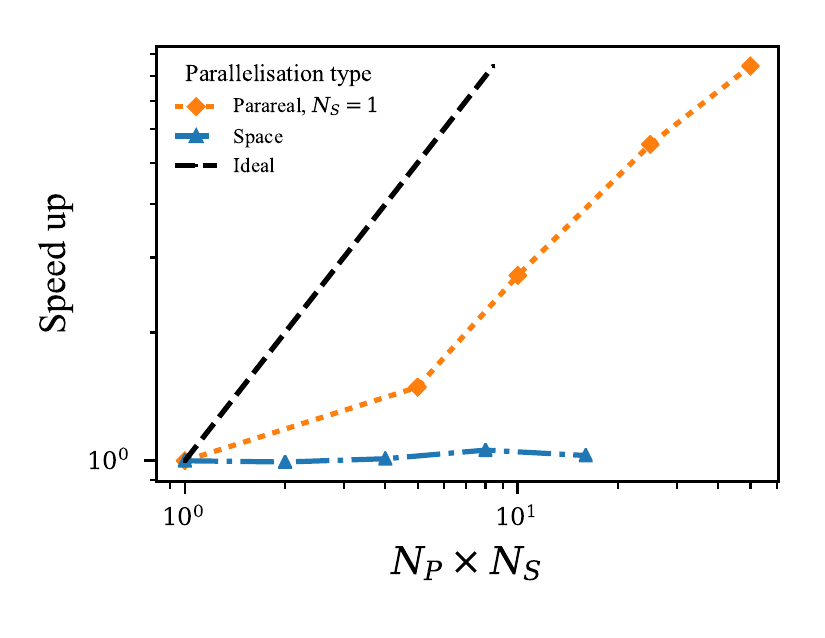}
\captionsetup{width=0.75\linewidth}
\caption{$R_m=3$, Speed up}
\end{subfigure}%
\begin{subfigure}{0.49\linewidth}
\includegraphics[width=\textwidth]{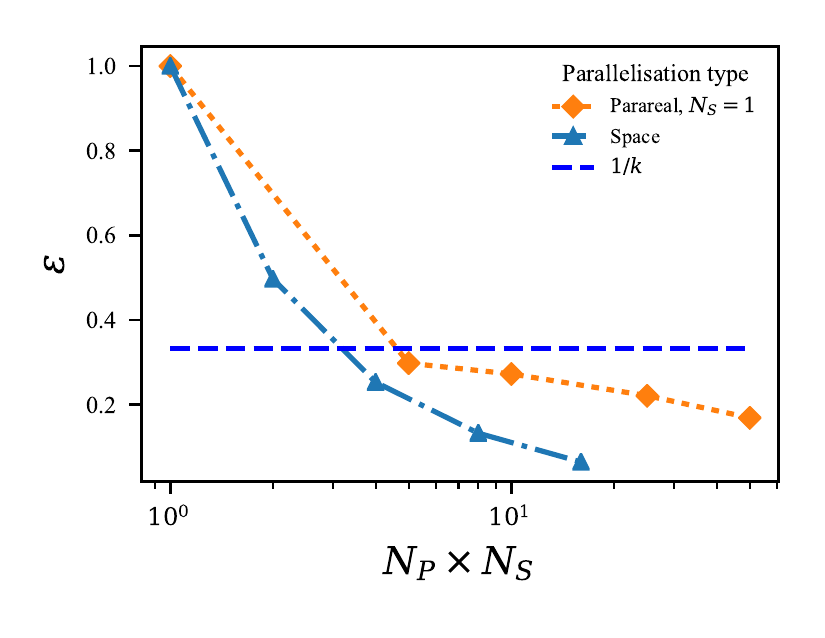}
\captionsetup{width=0.75\linewidth}
\caption{$R_m=3$, Efficiency}
\end{subfigure}
\begin{subfigure}{0.49\linewidth}
\includegraphics[width=\textwidth]{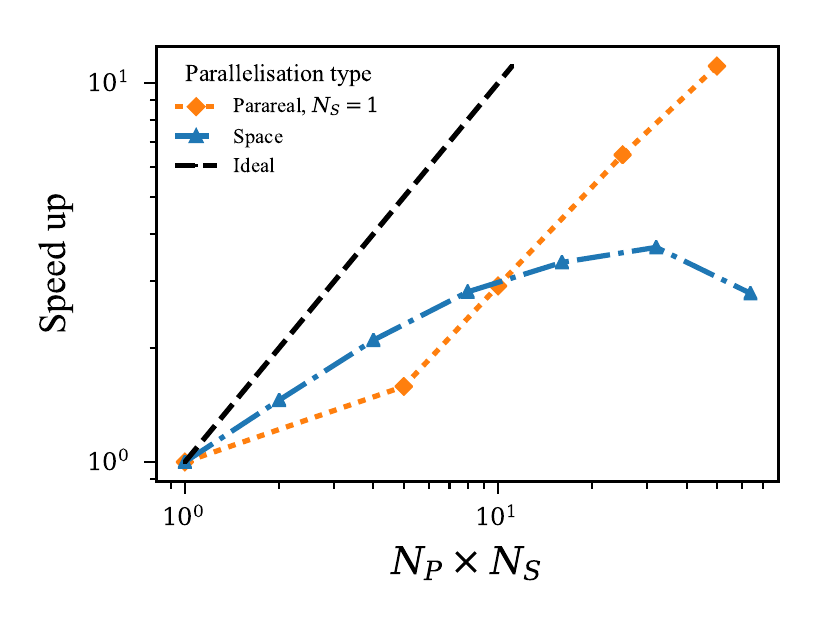}
\captionsetup{width=0.75\linewidth}
\caption{$R_m=300$, Speed up}
\end{subfigure}%
\begin{subfigure}{0.49\linewidth}
\includegraphics[width=\textwidth]{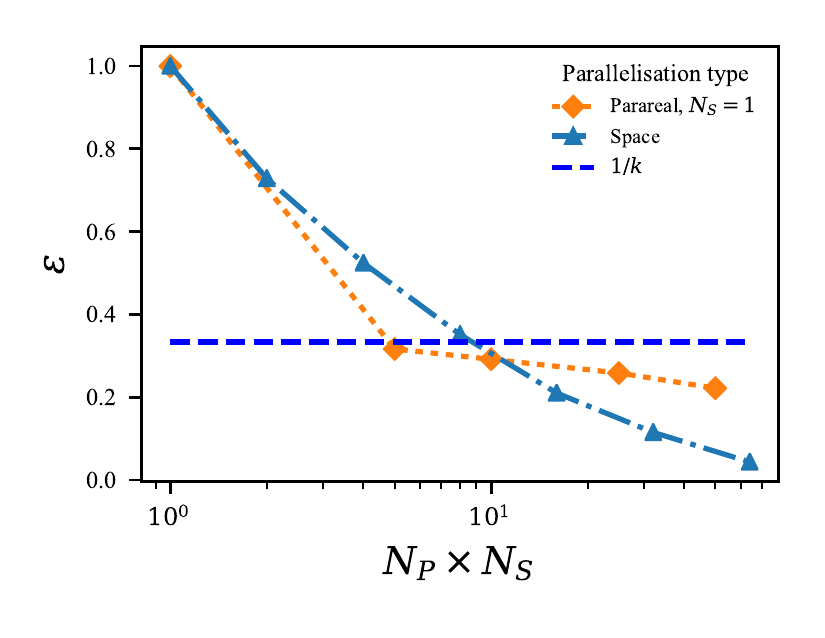}
\captionsetup{width=0.75\linewidth}
\caption{$R_m=300$, Efficiency}
\end{subfigure}
\begin{subfigure}{0.49\linewidth}
\includegraphics[width=\textwidth]{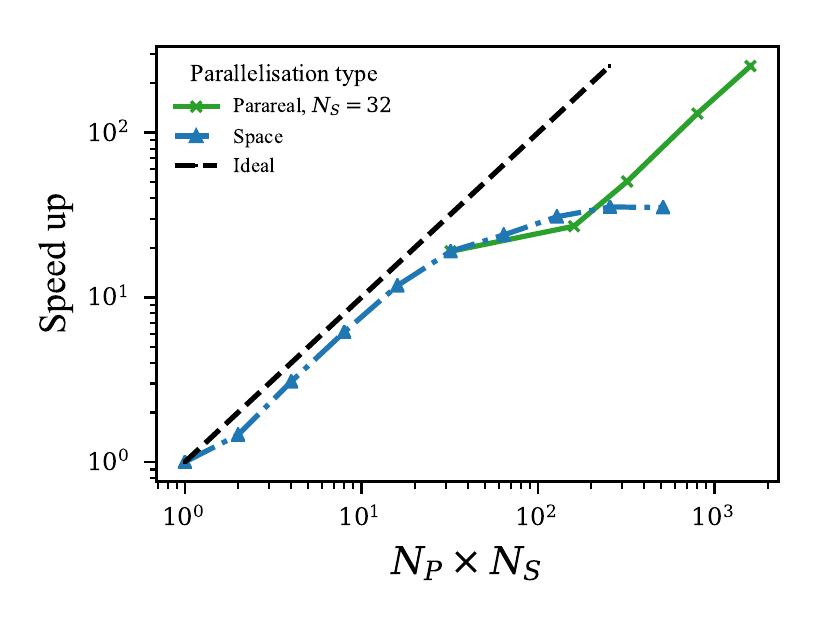}
\captionsetup{width=0.75\linewidth}
\caption{$R_m=3000$, Speed up}
\end{subfigure}%
\begin{subfigure}{0.49\linewidth}
\includegraphics[width=\textwidth]{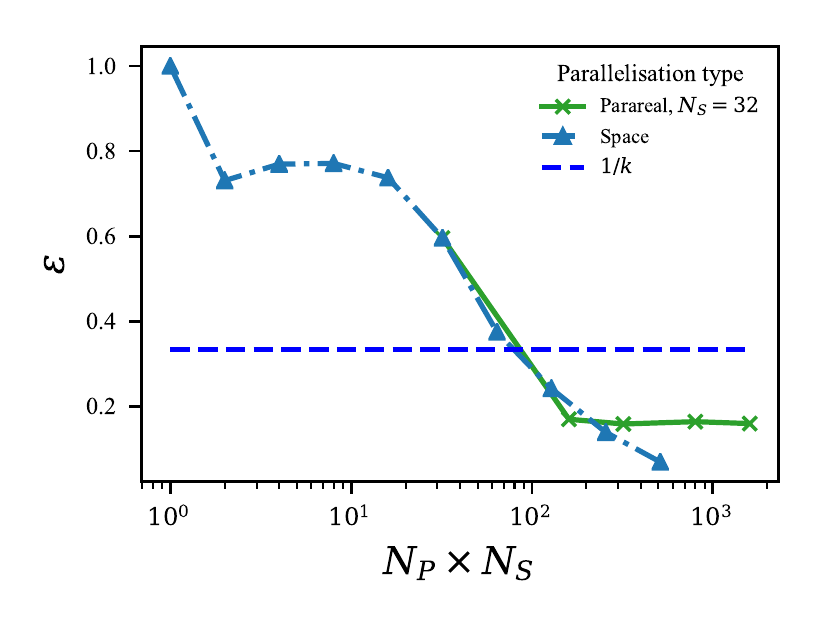}
\captionsetup{width=0.75\linewidth}
\caption{$R_m=3000$, Efficiency}
\end{subfigure}
\caption{Speed up and parallel efficiency of the parareal method compared to parallelisation in space for $R_m$ of 3, 300, and 3000, Galloway Proctor flow. Total number of processors is calculated as number of processors in space ($N_S$) multiplied by number of processors for Parareal ($N_P$). In the case of $R_m$=3000, parareal simulations were carried out with 32 processors in space, as serial runs with one processor were time intensive. Spatial resolutions required were $16^2$, $128^2$, and $512^2$ respectively. Results here are more promising than in the Roberts flow. Parareal becomes more efficient than spatial parallelisation for smaller processor numbers, and keeps higher efficiency for longer, closer to the theoretical maximum of $1/k$ ($k$: number of Parareal iterations). Scaling saturation for parareal has not been reached even at 1600 processors in the $R_m$=3000 case. }
\label{fig:rob_speed_up_efficiency_gp}
\end{figure}

\begin{figure}[ht]
\centering

\includegraphics[width=.5\linewidth]{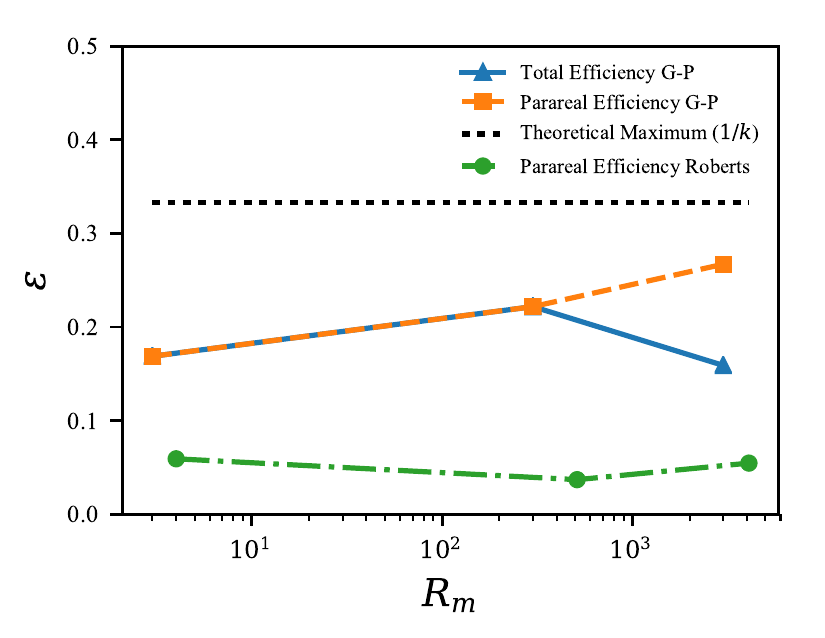}

\caption{Parallel efficiency vs $R_m$ for Galloway-Proctor and Roberts dynamos. Galloway-Proctor results show higher efficiency than the Roberts flow. Parallel efficiency of the method does not appear to degrade with increasing $R_m$. There is a reduction for total efficiency for $R_m$=3000, however, this is due to a combination of the efficiency of the spatial parallelisation with the parareal efficiency. Efficiency of parareal alone is comparable to the efficiency of the lower $R_m$ simulations in the Galloway-Proctor case.}
\label{fig:parareal_v_rm}
\end{figure}

Results for $Rm$= 3, 300 and 3000 for the Galloway Proctor flow are shown in Figure~\ref{fig:rob_speed_up_efficiency_gp}. 
Performance of Parareal is much better than for the Roberts flow. 
Parareal speed up is competitive with spatial parallelisation at a relatively low number of processors, and the efficiency of Parareal stays close to $1/3$ over a range of $N_p$ and does not fall much as the number of processors increases. 
The poor performance of Dedalus in parallelising the $Rm$=3 case is attributable to the fact that there are only $16^2$ spectral modes. 
In the $R_m=3000$ case, the results show parallel in space and time results, with 32 processors in space. 
The results show that speed up above that of spatial parallelisation alone is possible, with an efficiency around 0.16. 
In all cases, Parareal has not yet reached saturation in its scaling performance, and has almost ideal scaling behaviour, except for a constant offset due to the bounds on Parareal scaling. 
In all of these cases, the number of iterations required to converge was three, so that the efficiency is bounded by $1/3$. 
This is shown in Fig.~\ref{fig:parareal_v_rm}, where the efficiency of the method is tracked over the different $R_m$. 
Efficiency is close to the bound of $1/3$, and Parareal efficiency does not fall with increasing $R_m$. 
Pure Parareal efficiency was estimated in the $R_m=3000$ case by dividing by the efficiency of the spatial parallelisation found for that particular $N_S$ (32). 
In that case, total parallel efficiency is lower than for the other Galloway-Proctor cases due to the combination of the spatial and temporal parallelisation, and is approximately the product of the two, as expected. 
This reduction in overall efficiency is unavoidable, as parallel in space is more efficient than Parareal for lower numbers of processors, and Parareal only becomes competitive after spatial efficiency falls away. Also shown on this Figure are the efficiencies obtained at different $R_m$ for the Roberts flow, highlighting the difference in performance of the method for the two cases.


\section{Conclusions}\label{sec:conclusions}

The Parareal algorithm has been found to offer parallel speed up for kinematic dynamo simulations beyond what can be achieved through spatial parallelisation alone. 
In the case of the simpler steady Roberts dynamo, the speed up is modest and parallel efficiencies are low.
Here, owing to the steady nature of the imposed velocity, the difference in computational complexity of the coarse and fine methods was found to be too small for good performance of Parareal.
The issue was that the time step size was not a limiting factor on the accuracy of the fine solver; as long as the time step was stable, it was within the accuracy required. 
Therefore, there was little room to use a coarser resolution for the coarse propagator in Parareal.

Performance for the time-dependent Galloway-Proctor flow was better and the efficiency stayed close to the theoretical limit over a wide range of magnetic Reynolds numbers, while scaling well to large numbers of processors.
In this problem, since evolution of the magnetic field depends explicitly on the current time, the accuracy of the solution depends more on the size of the time step. 
This means that a time step in the coarse solver much larger than that of the fine step is possible, allowing for better Parareal performance.
Fully coupled dynamic dynamo simulation is complicated, and has non-linear dependencies, and so the accuracy of the fine solver is expected to behave more like the Galloway Proctor flow. Therefore, the good performance of Parareal for the Galloway-Proctor flow suggests that good performance can be expected also for more complex dynamos.
This paper therefore gives an interesting example of how studying a too simple problem can lead to an overly negative assessment of the performance of Parareal.

The parallel efficiency of the Parareal algorithm applied to the Galloway Proctor dynamo is close to the theoretical maximum of $1/k_{con}$. This means that the overheads due to communication are small, in comparison to the serial cost of the coarse method, pointing to an efficient implementation of the algorithm. 
Performance of the algorithm when applied to this problem does not appear to degrade with increased $R_m$, as can be seen in Figure~\ref{fig:parareal_v_rm}. 
Performance has remained constant, with Parareal efficiency not much lower than $1/3$ for $R_m= 3$, $R_m=300$ and $R_m=3000$. 
This is noteworthy since highly advective problems are thought to cause problems with Parareal convergence, but this has not yet been found in the the highly advective case with $R_m$ up to $\sim 10^3$. 

The results in this paper show that parallel in time methods can speed up dynamo simulations and are therefore worthy of further study.
Better parallelization would enable the study of dynamos at larger magnetic Reynolds numbers by reducing simulation times by harnessing the ever growing number of available cores in HPC facilities.
Current and future work involves extending our analysis to non-linear dynamic dynamo systems and the interaction of magnetic fields with convection. The development of successful parallel-in-time methods for these problems could allow the integration of geo- and astro- dynamo simulations in parameter regimes closer to reality than have been hitherto possible.  

\section*{Acknowledgments}
A.T.C is supported by the Engineering and Physical Sciences Research Council (EPSRC) Centre for Doctoral Training in Fluid Dynamics (EP/L01615X/1). 
C.J.D is supported by a Natural Environment Research Council (NERC) Independent Research Fellowship (NE/L011328/1). 
D. R. thankfully acknowledges support from grants EPSRC EP/P02372X/1 "A new algorithm to track fast ions in fusion reactors" and NERC NE/R008795/1 "Parallel Paradigms for Numerical Weather Prediction".
S.M.T is supported by a Levehulme Fellowship and by funding from the European Research Council (ERC) under the European Union's Horizon 2020 research and innovation programme (grant agreement no. D5S-DLV-786780).
This work used the ARCHER UK National Supercomputing Service (http://www.archer.ac.uk) as well as ARC2 and ARC3, part of the High Performance Computing facilities at the University of Leeds. 
Figures were produced using Matplotlib \cite{hunter2007matplotlib}. All codes used to create the results in this manuscript are publicly available on GitHub at \cite{andrewclarke_2019}.

\FloatBarrier

\bibliographystyle{model1-num-names}
\bibliography{lib_bibliography.bib}

\begin{thebibliography}{48}
\expandafter\ifx\csname natexlab\endcsname\relax\def\natexlab#1{#1}\fi
\providecommand{\url}[1]{\texttt{#1}}
\providecommand{\href}[2]{#2}
\providecommand{\path}[1]{#1}
\providecommand{\DOIprefix}{doi:}
\providecommand{\ArXivprefix}{arXiv:}
\providecommand{\URLprefix}{URL: }
\providecommand{\Pubmedprefix}{pmid:}
\providecommand{\doi}[1]{\href{http://dx.doi.org/#1}{\path{#1}}}
\providecommand{\Pubmed}[1]{\href{pmid:#1}{\path{#1}}}
\providecommand{\bibinfo}[2]{#2}
\ifx\xfnm\relax \def\xfnm[#1]{\unskip,\space#1}\fi
\bibitem[{Roberts and Soward(1992)}]{roberts1992dynamo}
\bibinfo{author}{P.~H. Roberts}, \bibinfo{author}{A.~M. Soward},
\newblock \bibinfo{title}{Dynamo theory},
\newblock \bibinfo{journal}{Annual review of fluid mechanics}
  \bibinfo{volume}{24} (\bibinfo{year}{1992}) \bibinfo{pages}{459--512}.
\bibitem[{Weiss(2002)}]{weiss2002dynamos}
\bibinfo{author}{N.~Weiss},
\newblock \bibinfo{title}{Dynamos in planets, stars and galaxies},
\newblock \bibinfo{journal}{Astronomy \& Geophysics} \bibinfo{volume}{43}
  (\bibinfo{year}{2002}) \bibinfo{pages}{3--9}.
\bibitem[{Moffatt(1978)}]{moffatt1978field}
\bibinfo{author}{H.~K. Moffatt},
\newblock \bibinfo{title}{Field generation in electrically conducting fluids},
\newblock \bibinfo{journal}{Cambridge University Press, Cambridge, London, New
  York, Melbourne}  (\bibinfo{year}{1978}).
\bibitem[{Roberts(1967)}]{roberts1967introduction}
\bibinfo{author}{P.~H. Roberts}, \bibinfo{title}{An introduction to
  magnetohydrodynamics}, volume~\bibinfo{volume}{6},
  \bibinfo{publisher}{Longmans London}, \bibinfo{year}{1967}.
\bibitem[{Davies et~al.(2011)Davies, Gubbins, and
  Jimack}]{davies2011scalability}
\bibinfo{author}{C.~J. Davies}, \bibinfo{author}{D.~Gubbins},
  \bibinfo{author}{P.~K. Jimack},
\newblock \bibinfo{title}{Scalability of pseudospectral methods for geodynamo
  simulations},
\newblock \bibinfo{journal}{Concurrency and Computation: Practice and
  Experience} \bibinfo{volume}{23} (\bibinfo{year}{2011})
  \bibinfo{pages}{38--56}.
\bibitem[{Knaepen and Moreau(2008)}]{knaepen2008magnetohydrodynamic}
\bibinfo{author}{B.~Knaepen}, \bibinfo{author}{R.~Moreau},
\newblock \bibinfo{title}{Magnetohydrodynamic turbulence at low magnetic
  reynolds number},
\newblock \bibinfo{journal}{Annu. Rev. Fluid Mech.} \bibinfo{volume}{40}
  (\bibinfo{year}{2008}) \bibinfo{pages}{25--45}.
\bibitem[{Kono and Roberts(2002)}]{kono2002recent}
\bibinfo{author}{M.~Kono}, \bibinfo{author}{P.~H. Roberts},
\newblock \bibinfo{title}{Recent geodynamo simulations and observations of the
  geomagnetic field},
\newblock \bibinfo{journal}{Reviews of Geophysics} \bibinfo{volume}{40}
  (\bibinfo{year}{2002}) \bibinfo{pages}{4--1}.
\bibitem[{Ossendrijver(2003)}]{Ossendrijver2003}
\bibinfo{author}{M.~Ossendrijver},
\newblock \bibinfo{title}{The solar dynamo},
\newblock \bibinfo{journal}{The Astronomy and Astrophysics Review}
  \bibinfo{volume}{11} (\bibinfo{year}{2003}) \bibinfo{pages}{287--367}.
\bibitem[{Roberts(1972)}]{roberts1972dynamo}
\bibinfo{author}{G.~O. Roberts},
\newblock \bibinfo{title}{Dynamo action of fluid motions with two-dimensional
  periodicity},
\newblock \bibinfo{journal}{Phil. Trans. R. Soc. Lond. A} \bibinfo{volume}{271}
  (\bibinfo{year}{1972}) \bibinfo{pages}{411--454}.
\bibitem[{Galloway and Proctor(1992)}]{galloway1992numerical}
\bibinfo{author}{D.~J. Galloway}, \bibinfo{author}{M.~R. Proctor},
\newblock \bibinfo{title}{Numerical calculations of fast dynamos in smooth
  velocity fields with realistic diffusion},
\newblock \bibinfo{journal}{Nature} \bibinfo{volume}{356}
  (\bibinfo{year}{1992}) \bibinfo{pages}{691}.
\bibitem[{Plunian and Radler(2009)}]{plunian2009harmonic}
\bibinfo{author}{F.~Plunian}, \bibinfo{author}{K.-H. Radler},
\newblock \bibinfo{title}{Harmonic and subharmonic solutions of the roberts
  dynamo problem. application to the karlsruhe experiment},
\newblock \bibinfo{journal}{arXiv preprint arXiv:0905.0847}
  (\bibinfo{year}{2009}).
\bibitem[{Tobias and Cattaneo(2013)}]{tobias2013shear}
\bibinfo{author}{S.~M. Tobias}, \bibinfo{author}{F.~Cattaneo},
\newblock \bibinfo{title}{Shear-driven dynamo waves at high magnetic reynolds
  number},
\newblock \bibinfo{journal}{Nature} \bibinfo{volume}{497}
  (\bibinfo{year}{2013}) \bibinfo{pages}{463--465}.
\bibitem[{Jones(2008)}]{jones2008course}
\bibinfo{author}{C.~A. Jones},
\newblock \bibinfo{title}{Course 2 dynamo theory},
\newblock \bibinfo{journal}{Les Houches} \bibinfo{volume}{88}
  (\bibinfo{year}{2008}) \bibinfo{pages}{45--135}.
\bibitem[{Matsui et~al.(2016)Matsui, Heien, Aubert, Aurnou, Avery, Brown,
  Buffett, Busse, Christensen, Davies et~al.}]{matsui2016performance}
\bibinfo{author}{H.~Matsui}, \bibinfo{author}{E.~Heien},
  \bibinfo{author}{J.~Aubert}, \bibinfo{author}{J.~M. Aurnou},
  \bibinfo{author}{M.~Avery}, \bibinfo{author}{B.~Brown},
  \bibinfo{author}{B.~A. Buffett}, \bibinfo{author}{F.~Busse},
  \bibinfo{author}{U.~R. Christensen}, \bibinfo{author}{C.~J. Davies}, et~al.,
\newblock \bibinfo{title}{Performance benchmarks for a next generation
  numerical dynamo model},
\newblock \bibinfo{journal}{Geochemistry, Geophysics, Geosystems}
  \bibinfo{volume}{17} (\bibinfo{year}{2016}) \bibinfo{pages}{1586--1607}.
\bibitem[{Schaeffer et~al.(2017)Schaeffer, Jault, Nataf, and
  Fournier}]{schaeffer2017turbulent}
\bibinfo{author}{N.~Schaeffer}, \bibinfo{author}{D.~Jault},
  \bibinfo{author}{H.-C. Nataf}, \bibinfo{author}{A.~Fournier},
\newblock \bibinfo{title}{Turbulent geodynamo simulations: a leap towards
  earth’s core},
\newblock \bibinfo{journal}{Geophysical Journal International}
  \bibinfo{volume}{211} (\bibinfo{year}{2017}) \bibinfo{pages}{1--29}.
\bibitem[{Mininni et~al.(2011)Mininni, Rosenberg, Reddy, and
  Pouquet}]{mininni2011hybrid}
\bibinfo{author}{P.~D. Mininni}, \bibinfo{author}{D.~Rosenberg},
  \bibinfo{author}{R.~Reddy}, \bibinfo{author}{A.~Pouquet},
\newblock \bibinfo{title}{A hybrid mpi--openmp scheme for scalable parallel
  pseudospectral computations for fluid turbulence},
\newblock \bibinfo{journal}{Parallel Computing} \bibinfo{volume}{37}
  (\bibinfo{year}{2011}) \bibinfo{pages}{316--326}.
\bibitem[{Croce et~al.(2014)Croce, Ruprecht, and Krause}]{croce2014parallel}
\bibinfo{author}{R.~Croce}, \bibinfo{author}{D.~Ruprecht},
  \bibinfo{author}{R.~Krause},
\newblock \bibinfo{title}{Parallel-in-space-and-time simulation of the
  three-dimensional, unsteady navier-stokes equations for incompressible flow},
\newblock in: \bibinfo{booktitle}{Modeling, Simulation and Optimization of
  Complex Processes-HPSC 2012}, \bibinfo{publisher}{Springer},
  \bibinfo{year}{2014}, pp. \bibinfo{pages}{13--23}.
\bibitem[{Gander(2015)}]{gander201550}
\bibinfo{author}{M.~J. Gander},
\newblock \bibinfo{title}{50 years of time parallel time integration},
\newblock \bibinfo{journal}{Multiple Shooting and Time Domain Decomposition
  Methods: MuS-TDD, Heidelberg, May 6-8, 2013} \bibinfo{volume}{9}
  (\bibinfo{year}{2015}) \bibinfo{pages}{69}.
\bibitem[{Lions et~al.(2001)Lions, Maday, and Turinici}]{lions2001resolution}
\bibinfo{author}{J.-L. Lions}, \bibinfo{author}{Y.~Maday},
  \bibinfo{author}{G.~Turinici},
\newblock \bibinfo{title}{R{\'e}solution d'edp par un sch{\'e}ma en temps
  parar{\'e}el},
\newblock \bibinfo{journal}{Comptes Rendus de l'Acad{\'e}mie des
  Sciences-Series I-Mathematics} \bibinfo{volume}{332} (\bibinfo{year}{2001})
  \bibinfo{pages}{661--668}.
\bibitem[{Minion(2010)}]{minion2010hybrid}
\bibinfo{author}{M.~L. Minion},
\newblock \bibinfo{title}{{A Hybrid Parareal Spectral Deferred Corrections
  Method}},
\newblock \bibinfo{journal}{Communications in Applied Mathematics and
  Computational Science} \bibinfo{volume}{5} (\bibinfo{year}{2010})
  \bibinfo{pages}{265--301}.
\bibitem[{Cortial and Farhat(2009)}]{CortialFarhat2009}
\bibinfo{author}{J.~Cortial}, \bibinfo{author}{C.~Farhat},
\newblock \bibinfo{title}{{A time-parallel implicit method for accelerating the
  solution of non-linear structural dynamics problems}},
\newblock \bibinfo{journal}{International Journal for Numerical Methods in
  Engineering} \bibinfo{volume}{77} (\bibinfo{year}{2009})
  \bibinfo{pages}{451--470}.
\bibitem[{Gander and Guttel(2013)}]{gander2013paraexp}
\bibinfo{author}{M.~J. Gander}, \bibinfo{author}{S.~Guttel},
\newblock \bibinfo{title}{Paraexp: A parallel integrator for linear
  initial-value problems},
\newblock \bibinfo{journal}{SIAM Journal on Scientific Computing}
  \bibinfo{volume}{35} (\bibinfo{year}{2013}) \bibinfo{pages}{C123--C142}.
\bibitem[{Fischer et~al.(2005)Fischer, Hecht, and Maday}]{fischer2005parareal}
\bibinfo{author}{P.~F. Fischer}, \bibinfo{author}{F.~Hecht},
  \bibinfo{author}{Y.~Maday},
\newblock \bibinfo{title}{A parareal in time semi-implicit approximation of the
  navier-stokes equations},
\newblock in: \bibinfo{booktitle}{Domain decomposition methods in science and
  engineering}, \bibinfo{publisher}{Springer}, \bibinfo{year}{2005}, pp.
  \bibinfo{pages}{433--440}.
\bibitem[{Samaddar et~al.(2017)Samaddar, Coster, Bonnin, Bergmeister,
  Havl{\'\i}c̆kov{\'a}, Berry, Elwasif, and Batchelor}]{samaddar2017temporal}
\bibinfo{author}{D.~Samaddar}, \bibinfo{author}{D.~Coster},
  \bibinfo{author}{X.~Bonnin}, \bibinfo{author}{C.~Bergmeister},
  \bibinfo{author}{E.~Havl{\'\i}c̆kov{\'a}}, \bibinfo{author}{L.~A. Berry},
  \bibinfo{author}{W.~R. Elwasif}, \bibinfo{author}{D.~B. Batchelor},
\newblock \bibinfo{title}{Temporal parallelization of edge plasma simulations
  using the parareal algorithm and the solps code},
\newblock \bibinfo{journal}{Computer Physics Communications}
  \bibinfo{volume}{221} (\bibinfo{year}{2017}) \bibinfo{pages}{19--27}.
\bibitem[{Bal and Maday(2002)}]{bal2002parareal}
\bibinfo{author}{G.~Bal}, \bibinfo{author}{Y.~Maday},
\newblock \bibinfo{title}{A “parareal” time discretization for non-linear
  pde’s with application to the pricing of an american put},
\newblock in: \bibinfo{booktitle}{Recent developments in domain decomposition
  methods}, \bibinfo{publisher}{Springer}, \bibinfo{year}{2002}, pp.
  \bibinfo{pages}{189--202}.
\bibitem[{Samuel(2012)}]{samuel2012time}
\bibinfo{author}{H.~Samuel},
\newblock \bibinfo{title}{Time domain parallelization for computational
  geodynamics},
\newblock \bibinfo{journal}{Geochemistry, Geophysics, Geosystems}
  \bibinfo{volume}{13} (\bibinfo{year}{2012}).
\bibitem[{Gander and Vandewalle(2007)}]{gander2007analysis}
\bibinfo{author}{M.~J. Gander}, \bibinfo{author}{S.~Vandewalle},
\newblock \bibinfo{title}{Analysis of the parareal time-parallel
  time-integration method},
\newblock \bibinfo{journal}{SIAM Journal on Scientific Computing}
  \bibinfo{volume}{29} (\bibinfo{year}{2007}) \bibinfo{pages}{556--578}.
\bibitem[{Steiner et~al.(2015)Steiner, Ruprecht, Speck, and
  Krause}]{steiner2015convergence}
\bibinfo{author}{J.~Steiner}, \bibinfo{author}{D.~Ruprecht},
  \bibinfo{author}{R.~Speck}, \bibinfo{author}{R.~Krause},
\newblock \bibinfo{title}{Convergence of parareal for the navier-stokes
  equations depending on the reynolds number},
\newblock in: \bibinfo{booktitle}{Numerical Mathematics and Advanced
  Applications-ENUMATH 2013}, \bibinfo{publisher}{Springer},
  \bibinfo{year}{2015}, pp. \bibinfo{pages}{195--202}.
\bibitem[{Ruprecht(2018)}]{Ruprecht2018}
\bibinfo{author}{D.~Ruprecht},
\newblock \bibinfo{title}{Wave propagation characteristics of parareal},
\newblock \bibinfo{journal}{Computing and Visualization in Science}
  \bibinfo{volume}{19} (\bibinfo{year}{2018}) \bibinfo{pages}{1--17}.
\bibitem[{Dai and Maday(2013)}]{dai2013stable}
\bibinfo{author}{X.~Dai}, \bibinfo{author}{Y.~Maday},
\newblock \bibinfo{title}{Stable parareal in time method for first-and
  second-order hyperbolic systems},
\newblock \bibinfo{journal}{SIAM Journal on Scientific Computing}
  \bibinfo{volume}{35} (\bibinfo{year}{2013}) \bibinfo{pages}{A52--A78}.
\bibitem[{Burns et~al.(2016)Burns, Vasil, Oishi, Lecoanet, and
  Brown}]{burns2016dedalus}
\bibinfo{author}{K.~J. Burns}, \bibinfo{author}{G.~M. Vasil},
  \bibinfo{author}{J.~S. Oishi}, \bibinfo{author}{D.~Lecoanet},
  \bibinfo{author}{B.~Brown}, \bibinfo{title}{Dedalus: Flexible framework for
  spectrally solving differential equations},
  \bibinfo{howpublished}{http://adsabs.harvard.edu/abs/2016ascl.soft03015B},
  \bibinfo{year}{2016}. \URLprefix \url{http://dedalus-project.org}.
\bibitem[{Galloway and Frisch(1984)}]{galloway1984numerical}
\bibinfo{author}{D.~Galloway}, \bibinfo{author}{U.~Frisch},
\newblock \bibinfo{title}{A numerical investigation of magnetic field
  generation in a flow with chaotic streamlines},
\newblock \bibinfo{journal}{Geophysical \& Astrophysical Fluid Dynamics}
  \bibinfo{volume}{29} (\bibinfo{year}{1984}) \bibinfo{pages}{13--18}.
\bibitem[{Aubanel(2011)}]{aubanel2011scheduling}
\bibinfo{author}{E.~Aubanel},
\newblock \bibinfo{title}{Scheduling of tasks in the parareal algorithm},
\newblock \bibinfo{journal}{Parallel Computing} \bibinfo{volume}{37}
  (\bibinfo{year}{2011}) \bibinfo{pages}{172--182}.
\bibitem[{Blouza et~al.(2009)Blouza, Laurent, and Kaber}]{blouzaetal2009}
\bibinfo{author}{A.~Blouza}, \bibinfo{author}{B.~Laurent},
  \bibinfo{author}{S.~M. Kaber},
\newblock \bibinfo{title}{{Parallel in time algorithms with reduction methods
  for solving chemical kinetics}},
\newblock \bibinfo{journal}{Communications in Applied Mathematics and
  Computational Science} \bibinfo{volume}{5} (\bibinfo{year}{2009})
  \bibinfo{pages}{241--263}.
\bibitem[{Baffico et~al.(2002)Baffico, Bernard, Maday, Turinici, and
  Z{\'e}rah}]{baffico2002parallel}
\bibinfo{author}{L.~Baffico}, \bibinfo{author}{S.~Bernard},
  \bibinfo{author}{Y.~Maday}, \bibinfo{author}{G.~Turinici},
  \bibinfo{author}{G.~Z{\'e}rah},
\newblock \bibinfo{title}{Parallel-in-time molecular-dynamics simulations},
\newblock \bibinfo{journal}{Physical Review E} \bibinfo{volume}{66}
  (\bibinfo{year}{2002}) \bibinfo{pages}{057701}.
\bibitem[{Maday and Turinici(2003)}]{maday2003parallel}
\bibinfo{author}{Y.~Maday}, \bibinfo{author}{G.~Turinici},
\newblock \bibinfo{title}{Parallel in time algorithms for quantum control:
  Parareal time discretization scheme},
\newblock \bibinfo{journal}{International journal of quantum chemistry}
  \bibinfo{volume}{93} (\bibinfo{year}{2003}) \bibinfo{pages}{223--228}.
\bibitem[{Maday(2007)}]{maday2007parareal}
\bibinfo{author}{Y.~Maday},
\newblock \bibinfo{title}{Parareal in time algorithm for kinetic systems based
  on model reduction},
\newblock \bibinfo{journal}{High-dimensional partial differential equations in
  science and engineering} \bibinfo{volume}{41} (\bibinfo{year}{2007})
  \bibinfo{pages}{183--194}.
\bibitem[{Lunet et~al.(2018)Lunet, Bodart, Gratton, and
  Vasseur}]{lunet2018time}
\bibinfo{author}{T.~Lunet}, \bibinfo{author}{J.~Bodart},
  \bibinfo{author}{S.~Gratton}, \bibinfo{author}{X.~Vasseur},
\newblock \bibinfo{title}{Time-parallel simulation of the decay of homogeneous
  turbulence using parareal with spatial coarsening},
\newblock \bibinfo{journal}{Computing and Visualization in Science}
  \bibinfo{volume}{19} (\bibinfo{year}{2018}) \bibinfo{pages}{31--44}.
\bibitem[{Ruprecht(2014)}]{ruprecht2014convergence}
\bibinfo{author}{D.~Ruprecht},
\newblock \bibinfo{title}{Convergence of parareal with spatial coarsening},
\newblock \bibinfo{journal}{PAMM} \bibinfo{volume}{14} (\bibinfo{year}{2014})
  \bibinfo{pages}{1031--1034}.
\bibitem[{Burns et~al.(tion)Burns, Vasil, Oishi, Lecoanet, Brown, and
  Quataert}]{burns-prep-dedalus}
\bibinfo{author}{K.~Burns}, \bibinfo{author}{G.~Vasil},
  \bibinfo{author}{J.~Oishi}, \bibinfo{author}{D.~Lecoanet},
  \bibinfo{author}{B.~Brown}, \bibinfo{author}{E.~Quataert},
\newblock \bibinfo{title}{Dedalus: A flexible framework for spectrally solving
  differential equations}  (\bibinfo{year}{In preparation}).
\bibitem[{Dalc{\'\i}n et~al.(2005)Dalc{\'\i}n, Paz, and Storti}]{dalcin2005mpi}
\bibinfo{author}{L.~Dalc{\'\i}n}, \bibinfo{author}{R.~Paz},
  \bibinfo{author}{M.~Storti},
\newblock \bibinfo{title}{Mpi for python},
\newblock \bibinfo{journal}{Journal of Parallel and Distributed Computing}
  \bibinfo{volume}{65} (\bibinfo{year}{2005}) \bibinfo{pages}{1108--1115}.
\bibitem[{Ascher et~al.(1997)Ascher, Ruuth, and Spiteri}]{ascher1997implicit}
\bibinfo{author}{U.~M. Ascher}, \bibinfo{author}{S.~J. Ruuth},
  \bibinfo{author}{R.~J. Spiteri},
\newblock \bibinfo{title}{Implicit-explicit runge-kutta methods for
  time-dependent partial differential equations},
\newblock \bibinfo{journal}{Applied Numerical Mathematics} \bibinfo{volume}{25}
  (\bibinfo{year}{1997}) \bibinfo{pages}{151--167}.
\bibitem[{Spalart et~al.(1991)Spalart, Moser, and Rogers}]{spalart1991spectral}
\bibinfo{author}{P.~R. Spalart}, \bibinfo{author}{R.~D. Moser},
  \bibinfo{author}{M.~M. Rogers},
\newblock \bibinfo{title}{Spectral methods for the navier-stokes equations with
  one infinite and two periodic directions},
\newblock \bibinfo{journal}{Journal of Computational Physics}
  \bibinfo{volume}{96} (\bibinfo{year}{1991}) \bibinfo{pages}{297--324}.
\bibitem[{Wang and Ruuth(2008)}]{wang2008variable}
\bibinfo{author}{D.~Wang}, \bibinfo{author}{S.~J. Ruuth},
\newblock \bibinfo{title}{Variable step-size implicit-explicit linear multistep
  methods for time-dependent partial differential equations},
\newblock \bibinfo{journal}{Journal of Computational Mathematics}
  (\bibinfo{year}{2008}) \bibinfo{pages}{838--855}.
\bibitem[{Smith and Tobias(2004)}]{smith2004vortex}
\bibinfo{author}{S.~G.~L. Smith}, \bibinfo{author}{S.~Tobias},
\newblock \bibinfo{title}{Vortex dynamos},
\newblock \bibinfo{journal}{Journal of Fluid Mechanics} \bibinfo{volume}{498}
  (\bibinfo{year}{2004}) \bibinfo{pages}{1--21}.
\bibitem[{Charbonneau(2012)}]{charbonneau2012solar}
\bibinfo{author}{P.~Charbonneau}, \bibinfo{title}{Solar and Stellar Dynamos:
  Saas-Fee Advanced Course 39 Swiss Society for Astrophysics and Astronomy},
  volume~\bibinfo{volume}{39}, \bibinfo{publisher}{Springer Science \& Business
  Media}, \bibinfo{year}{2012}.
\bibitem[{Hunter(2007)}]{hunter2007matplotlib}
\bibinfo{author}{J.~D. Hunter},
\newblock \bibinfo{title}{Matplotlib: A 2d graphics environment},
\newblock \bibinfo{journal}{Computing in science \& engineering}
  \bibinfo{volume}{9} (\bibinfo{year}{2007}) \bibinfo{pages}{90--95}.
\bibitem[{Clarke(2019)}]{andrewclarke_2019}
\bibinfo{author}{A.~Clarke}, \bibinfo{title}{Parareal-dynamo},
  \bibinfo{year}{2019}. \DOIprefix\doi{10.5281/zenodo.2554026}.

\end{thebibliography}

\end{document}